%% file: Eichhorn_et_al_2008_arXiv.tex
\documentclass{article}

\usepackage[pdftex]{graphicx}
\usepackage[square]{natbib}

\usepackage[sans]{dsfont}
\usepackage{url,amsmath,amsopn}
\usepackage[psamsfonts]{amssymb}
\usepackage{stmaryrd}
\usepackage{float}

\usepackage{fancyhdr}
\newcommand{\myem}{\em}
\renewcommand{\url}[1]{{\tt #1}}

\input{./definitions}
\graphicspath{{figures/}}

\begin{document}

\title{Natural Image Coding in V1: How Much Use is Orientation Selectivity?} 

\author{Jan Eichhorn, Fabian Sinz and Matthias
  Bethge\footnote{Corresponding Author}\\
Max Planck Institute for Biological Cybernetics\\
T{\"u}bingen, Germany}
\maketitle 

\vspace{-.5cm}

\begin{abstract}
  Orientation selectivity is the most striking feature of simple cell
  coding in V1 which has been shown to emerge from the reduction of
  higher-order correlations in natural images in a large variety of
  statistical image models.  The most parsimonious one among these
  models is linear Independent Component Analysis (ICA), whereas
  second-order decorrelation transformations such as Principal
  Component Analysis (PCA) do not yield oriented filters.  Because of
  this finding it has been suggested that the emergence of orientation
  selectivity may be explained by higher-order redundancy reduction.
  In order to assess the tenability of this hypothesis, it is an
  important empirical question how much more redundancies can be
  removed with ICA in comparison to PCA, or other second-order
  decorrelation methods. This question has not yet been settled, as
  over the last ten years contradicting results have been reported
  ranging from less than five to more than hundred percent extra gain
  for ICA. Here, we aim at resolving this conflict
  by presenting a very careful and comprehensive analysis using three
  evaluation criteria related to redundancy reduction: In addition to
  the multi-information and the average log-loss we compute, for the
  first time, complete rate-distortion curves for ICA in comparison
  with PCA. Without exception, we find that the advantage of the ICA
  filters is surprisingly small.  We conclude that orientation
  selective receptive fields in primary visual cortex cannot be
  explained in the framework of linear redundancy reduction.
  Furthermore, we show that a simple spherically
  symmetric distribution with only two parameters can fit the data
  even better than the probabilistic model underlying ICA. Since
  spherically symmetric models are agnostic with respect to the
  specific filter shapes, we conlude that 
  orientation selectivity is unlikely to play a critical role for redundancy
  reduction even if the class of transformations is not limited to
  linear ones.
\end{abstract}

\section*{Author Summary}
 What is the computational goal underlying the response properties of
 simple cells in primary visual cortex? Since the Nobel prize winning
 work of Hubel and Wiesel we have known that orientation selectivity is an
 important feature of simple cells in V1.  A conclusive and
 quantitative explanation for the actual purpose of this biological
 structure, however, still remains elusive. Among the hypotheses
on its purpose, the idea of redundancy reduction (Barlow, 1959)
stands out in being formalized precisely in terms of a quantitative
objective function.  While it has been reported previously that
orientation selectivity can reduce the redundancy in the
representation of color images by more than 100\%, we here provide
strong evidence that the actual gain is only 3\% for both color and
gray value images. Thus, our results challenge the conception that
the major goal of orientation selective coding in V1 is redundancy
reduction. Furthermore, we now have a reliable reference point for
quantitative investigations of nonlinear early vision models.

\section*{Introduction}

It is a long standing hypothesis that neural representations in
sensory systems are adapted to the statistical regularities of the
environment \cite{attneave:1954,barlow:1959}. Despite widespread
agreement that neural processing in the early visual system must be
influenced by the statistics of natural images, there are many
different viewpoints on how to precisely formulate the computational
goal the system is trying to achieve. At the same time, different
goals might be achieved by the same optimization criterion or learning
principle. Redundancy reduction \cite{barlow:1959}, the most prominent
example of such a principle, can be beneficial in various ways: it can
help to maximize the information to be sent through a channel of
limited capacity \cite{atick:1992}, it can be used to learn the
statistics of the input \cite{barlow:1989} or to facilitate pattern
recognition \cite{watanabe:1981}.

Besides redundancy reduction, a variety of other interesting
criteria such as {\em sparseness}
\cite{foldiak:1990,olshausen:1996}, {\em temporal coherence}
\cite{foldiak:1991}, or {\em predictive information}
\cite{bialek:2001,becker:1992} have been formulated. An important
commonality among all these ideas is the tight link to density
to density estimation of the input signal. Redundancy reduction, in
particular, can be interpreted as a special form of density estimation
where the goal is to find a mapping which transforms the data into a
representation with statistically independent coefficients. This idea
is related to projection pursuit \cite{Friedman:1984} where density
estimation is carried out by optimizing over a set of possible
transformations in order to match the statistics of the transformed
signal as good as possible to a pre-specified target distribution.
Once the distribution has been matched, applying the inverse
transformation effectively yields a density model for the original
data. From a neurobiological point of view, we may interpret the
neural response properties as an implementations of such transformations.
Accordingly, the redundancy reduction principle can be seen as special
case of projection pursuit where the target distribution is assumed to
have independent marginal distributions. This renders it a very
general principle. 

A crucial aspect of this kind of approach is the class of
transformations over which to optimize. From a statistician's point of
view it is important to choose a regularized function space in order
to avoid overfitting. On the other hand, if the class of possible
transformations is too restricted, it may be impossible to find a good
match to the target distribution. From a visual neuroscientist's point
of view, the choice of transformations should be related to the class
of possible computations in the early visual system.

Intriguingly, a number of response properties of visual neurons have
been reproduced by optimizing over the class of linear transformations
on natural images for redundancy reduction (for a
review see \cite{simoncelli:2001,li:2006}). For instance, Buchsbaum
and Gottschalk revealed a link between the second-order statistics of
color images and opponent color coding of retinal ganglion cells by
demonstrating that decorrelating natural images in the trichromatic
color space with Principal Component Analysis (PCA) yields the
luminance, the red-green, and the blue-yellow channel
\citep{buchsbaum:1983}.  Atick and Redlich derived the center-surround
receptive fields by optimizing a symmetric decorrelation
transformation \cite{atick:1992a}. Later, also spatio-temporal
correlations in natural images or sequences of natural images were
linked to the receptive field properties in the retina and the lateral
geniculate nucleus (LGN) \citep{vanHateren:1993,dong:1995,dan:1996}.

On the way from LGN to primary visual cortex, orientation selectivity
emerges as a striking new receptive field property. A number of
researchers \citep[e.g.][]{hancock:1992,li:1994} have used the
covariance properties of natural images to derive linear basis
functions that exhibit similar properties. Decorrelation alone,
however, was not sufficient to achieve this goal.  Rather, additional
constraints were necessary, such as spatial locality or symmetry.

It was not until the reduction of higher-order correlations were
taken into account that the derivation of localized and oriented
band-pass filters---resembling orientation selective receptive fields in V1---
was achieved without the necessity to assume any further constraints.
Those filters were derived with Independent Component Analysis (ICA),
a generalization of Principal Component Analysis (PCA), which in
addition seeks to reduce all higher-order correlations
\cite{olshausen:1996,bell:1997}.

This finding suggests that the orientation selectivity of V1 simple
cells may be {\it explained} by the minimization of higher-order
correlations. But in order to test this hypothesis, one must also
demonstrate that those filters can yield a substantial advantage
for redundancy reduction. If this is not the case, the objective of redundancy
reduction does not seem to be critical for the emergence of
orientation selectivity in primary visual cortex.

The importance of such a {\it quantitative} assessment has already
been pointed out in earlier studies \citep{lewicki:1999,wachtler:2001,
LeeWacSej2002,petrov:2003,bethge:2006,li:2006}. In particular,
a direct quantification of the redundancy reduction achieved with ICA
and PCA for natural images has been carried out in
\citep{lewicki:1999,wachtler:2001,LeeWacSej2002,bethge:2006}
but they differ considerably in their conclusions regarding the effect
of higher-order correlations in linear image models ranging from less
than five to more than hundred percent. 

This is a very unsatisfying situation as even for the simplest class
of models---the linear case---the literature does not provide a clear
answer to the question of how important orientation selectivity may be
for the reduction of redundancy in natural images. In order to resolve
this issue, we carried out this comprehensive study which 
exclusively investigates {\it linear} image models that have been
around for more than a decade by now. Our goal is to establish a
reliable reference against which more sophisticated image models can
be compared to in the future.  We elaborate on our own previous work
\citep{bethge:2006} by using an optimized ICA algorithm on the color
image data set for which the largest higher-order redundancy reduction
has been reported previously. The advantage of the resulting
orientation selective ICA filters is tested comprehensively with three
different types of analysis that are related to the notion of
redundancy reduction and coding efficiency: (A) multi-information
reduction, (B) average log-likelihood, and (C) rate-distortion curves.
Since we put strong emphasis on the reproducibility and verifiability
of our results, we provide our code and the dataset online ({\tt
http://www.kyb.tuebingen.mpg.de/bethge/code/QICA/}).

Our results show that orientation selective ICA filters do not excel
in any of these measures: We find that the gain of ICA in redundancy
reduction over a random decorrelation method is only about 3\% for
color and gray-value images.  In terms of rate-distortion curves, ICA
performs even worse than PCA. Furthermore, we demonstrate that a
simple spherically symmetric model with only two parameters fits the
data significantly better. Since in this model the specific shape of
the filters is ignored, we conclude that it is unlikely that orientation
selectivity plays a critical role for redundancy reduction even if the
class of transformations is not limited to linear ones.

\section*{Material and Methods}

An important difficulty in setting up a quantitative comparison
originates from the fact that it bears several issues that may be
critical for the results. In particular, choices have to be made
regarding the {\it evaluation criteria}, the {\it image data}, the
{\it estimation methods}, regarding which {\em linear transformations}
besides ICA to be included in the comparison, and which {\it
  particular implementation of ICA} to be used. The significance of
the outcome of the comparison will therefore depend on how careful
these choices have been made and the most relevant issues will be
addressed in the following.

\subsection*{Notation and Nomenclature}

For both, color and gray-value data, we write $\mathbf{x}$ to refer to
single vectors which contain the raw pixel intensities. Vectors are
indicated by bold font while the same letter in normal font with a
subindex denotes one of its components. Vectors without subindices
usually denote random variables, while subindices indicate specific
examples. In some cases it is convenient to define the corresponding
data matrix $X=(\mathbf{x}_1, \dots , \mathbf{x}_N)$ which holds
single images patches in its columns. The letter $N$ denotes the
number of examples in the dataset, while $n$ is used for the dimension
of a single data point.

Transformations are denoted by $W$, oftentimes with a subindex to
distinguish different types. The result of a transformation to either
a vector $\mathbf{x}$ or a data matrix $X$ will be written as
$\mathbf{y}=W \mathbf{x}$ or $Y=WX$, respectively.

Probability densities are denoted with the letters $p$ and $q$,
sometimes with a subindex to indicate differences between
distributions whenever it seems necessary for clarity. In general, we
use the hat symbol to distinguish between true entities and their
empirical estimates. For instance, $p_{\mathbf{y}}(\mathbf{y}) =
p_{\mathbf{x}}(W^{-1}\mathbf y)\cdot |\det W|^{-1}$ is the true
probability density of $\mathbf y$ after applying a fixed
transformation $W$, while $\hat p_{\mathbf y}(\mathbf y)$ refers to
the corresponding empirical estimate.

A distribution $p(\mathbf y)$ is called {\em factorial}, or {\em
  marginally independent}, if it can be written as a product of its
marginals, i.e. $p(\mathbf y) = \prod_{i=1}^n p_i(y_i)$ where $p_i(y_i)$
is obtained by integrating $p(\mathbf{y})$ over all components but
$y_i$.

Finally, the expectation over some entity $f$ with respect to $\mathbf
y$ is written as $\expc[\mathbf y]{f(\mathbf y)} = \int p(\mathbf{y})
f(\mathbf{y})d\mathbf{y}$. Sometimes, we use the density instead of
the random variable in the subindex to indicate the distribution, over
which the expectation is taken. If there is no risk for confusion we drop
the subindex. Just as above, the empirical expectation is marked with
a hat symbol, i.e. $\expe{f(\mathbf{y})}=\frac{1}{N} \sum_{k=1}^N
f(\mathbf{y}_k)$.

\subsection*{How to compare early vision models?}
\label{Sec:benchmarking}

A principal complicacy in low-level vision is the lack of a clearly
defined task.  Therefore, it is difficult to compare different image
representations as it is not obvious {\it a priori} what measure
should be used. 

\paragraph{(A) Multi-Information}

The first measure we consider is the {\em multi-information} \citep{perez:1977}, which
is the true objective function that is minimized by ICA over the choice of filters $W$. 
The multi-information assesses the
total amount of statistical dependencies between the components $y_i$
of a filtered patch $\mathbf{y}=W\mathbf{x}$:
\begin{equation}
  \label{eq:MultiInfo}
  I[p(\mathbf{y})]
  =
  D_\text{KL}\left[p(\mathbf{y})\left|{\atop}\!\!\right|\prod_j p_j(y_j)\right]
  = \expc[p]{\log \frac{p(\mathbf{y})}{\prod_j p_j(y_j)}}
  = \sum_{j=1}^n h[p_j(y_j)] - h[p(\mathbf y)].
\end{equation}
The terms $h[p_j(y_j)]$ and $h[p(\mathbf y)]$ denote the marginal and the joint
entropies of the true distribution, respectively. The
{\myem Kullback-Leibler-Divergence} or {\em Relative Entropy}
\[D_\text{KL}[p||q] = \expc[p]{\log\frac{p(\mathbf y)}{q(\mathbf
    y)}}\] is an information theoretic dissimilarity measure between
two distributions $p$ and $q$ \citep{cover:1991}. It is always
non-negative and zero if and only if $p$ equals $q$.  If the
redundancy reduction hypothesis is taken literally, the
multi-information is the right measure to minimize, since it measures
how close to factorial the true distribution of the image patches in
the representation $\mathbf y$ really is.

The application of linear ICA algorithms to ensembles of natural images
very reliably yields transformations consisting of localized and oriented
bandpass filters similar to the receptive fields of neurons in V1. It is less
clear, however, whether these filter properties also critical to the
minimization of the multi-information, i.e. the
redundancy?  In order to assess the tenability of the idea that a V1 simple cell
is adjusted to the purpose of redundancy reduction, it is important to
know whether such a tuning can---{\it in principle}---result in a large reduction of the
multi-information. One way to address this question is to measure {\it
  how much} more the multi-information is actually reduced by the ICA
filters in comparison to others such as PCA filters. This approach
has been taken in \citep{bethge:2006}.

One problem with estimating multi-information is that it involves the
joint entropy $h[p(\mathbf y)]$ of the true distribution which is
generally very hard to estimate. In certain cases, however,
the problem can be bypassed by evaluating the difference in the multi-information between two
representations $\mathbf x$ and $\mathbf y$. In particular, if $\mathbf{y}$ is related to $\mathbf{x}$ by the linear
transformation $\mathbf{y} = W \mathbf{x}$ it follows from
definition~\eqref{eq:MultiInfo} and the transformation theorem for
probability densities $$p_{\mathbf y}(\mathbf{y}) = p_{\mathbf x}(\mathbf{x}) \left |\det
  \left(\frac{\partial y}{\partial x}\right)\right|^{-1} = p_{\mathbf{x}}(W^{-1}\mathbf y)\cdot |\det W|^{-1}$$ that
difference in multi-information can be expressed as
\begin{eqnarray*}
  I[p(\mathbf{y})] - I[p(\mathbf{x})]
  &=& \sum_k h[p_k(y_k)] - h[p(y)] - \left(\sum_k h[p_k(x_k)] - h[p(x)]\right)\\
  &=& \sum_k h[p_k(y_k)] - \sum_k h[p_k(x_k)] - \log \left |\det W \right|\,.
\end{eqnarray*}
For
convenience, we chose a volume-conserving gauge \citep{bethge:2006}
where all linear decorrelation transforms are of determinant one, and 
hence $\log \left |\det W \right| = 0$. This means that differences in multi-information are equal to
differences of marginal entropies which can be estimated robustly. 
Thus, our empirical estimates of the multi-information differences are given by:
\begin{equation}\label{Eq:empirial_multi}
\Delta I \approx \sum_k h[\hat p_k(y_k)] - \sum_k h[\hat p_k(x_k)] \quad s.t. \,\, |\det(W)|=1
\end{equation}
For estimating the entropy of the univariate marginal distributions,
we employ the OPT estimator introduced in\citep{bethge:2006}
which uses the exponential power family to
fit the marginal distributions by searching for the OPTimal
exponent. This estimator has been shown to give highly reliable
results for natural images.  In particular, it is
much more robust than entropy estimators based on the sample
kurtosis which easily overestimate the multi-information.

\paragraph{(B) Average Log Loss (ALL)}

As mentioned earlier, redundancy reduction can be interpreted
as a special form of density estimation where the goal is to find a 
mapping which transforms the data into a representation with statistically independent 
coefficients. This means that any given transformation specifies a density model
over the data. Our second measure, the average log-loss (ALL), evaluates the agreement of this density model
with the actual distribution of the data:
\begin{equation}\label{eq:log_loss_def}
  \expc[p]{-\log \hat p(\mathbf{y})}= - \int p(\mathbf{y}) \log \hat
  p(\mathbf{y}) d\mathbf{y} = H[p] + D_\text{KL}[p||\hat p]
\end{equation}
The average log-loss is a principled measure quantifying how different the model density
$\hat p(\mathbf{y})$ is from the true density $p(\mathbf{y})$
\citep{bernardo:1979}.  Since the KL-divergence is positive and zero
if and only if $\hat p = p$ the ALL is minimal only if $\hat p$
matches the true density. Furthermore, differences in the average log-loss correspond to
differences in the coding cost (i.e.  information rate) in the case of
sufficiently fine quantization. For natural images, different image
representations have been compared with respect to this measure in
\citep{lewicki:1999,wachtler:2001,LeeWacSej2002}.  

For the estimation of the average log-loss, we follow Lewicki {\it et al.}
\citep{lewicki:1999,lewicki:2000} (We are referring here to
  the first method in the cited references. The defining equation
  there contains an extra term $N \log \sigma$ which indicates the
  role of noise. Note, however, that this term has only a symbolic
  meaning as the noise is assumed to be independent of the
  representation so that this term can in fact be ignored)  by simply
using the empirical average
\begin{equation}
  \label{eq:LogLoss}
  \expc[p]{-\log \hat p(\mathbf{y})}
  \, \approx \,
  \expe[\mathbf y]{-\log \hat p(\mathbf{y})}
  \,=\,
  - \frac{1}{N} \sum_{k=1}^N \log \hat p(\mathbf{y}_k) \,.
\end{equation}
While this estimator in principle can be prone to overfitting, we
control for this risk by evaluating all estimates on an independent
test set, whose data has not been used during the parameter fit.
Furthermore, we compare the average log-loss to the parametric
entropy estimates $h[\hat{p}]$ that we use in (A) for estimating
the multi-information changes (see Eq.~\ref{Eq:empirial_multi}).
The difference between both quantities has
been named {\it differential log-likelihood} \citep{vanHulle:2005} and
can be used to assess the goodness of fit of a model distribution:
\begin{eqnarray}
  \label{eq:DLL}
  \expe{-\log \hat{p}} - h[\hat{p}]
  &=& 
  \expc[\hat{p}]{\log \hat{p}} - \expe{\log \hat{p}} \nonumber 
  \,.
\end{eqnarray}

The shape of the parametric model is well matched to the actual
distribution if the differential log-likelihood converges to zero with
increasing number of data points.

\paragraph{(C) Rate-Distortion Curves} Finally, we consider {\it
  efficient coding} or {\it minimum mean square error reconstruction}
as a third objective. In contrast to the previous objectives, it is
now assumed that there is a bottleneck, limiting the amount of
information that can be transmitted, and the goal is to maximize the
amount of {\it relevant} information transmitted about the image.  In
the context of neural coding, the redundancy reduction hypothesis has
oftentimes been motivated in terms of coding efficiency. In fact,
instead of minimizing the multi-information one can equivalently ask
for the linear transformation $W$ which maximizes the mutual
information between its input $\mathbf{x}$ and its output
$W\mathbf{x}+\xi$ when additive noise $\xi$ is added to the output
\cite{nadal:1994,bell:1995}.  It is important to note, however, that
this minimalist approach of ``information maximization'' is ignorant
with respect to how useful or {\it relevant} the information is that
has been transmitted \cite{simoncelli:2001}.

For natural images, the source signal $\mathbf{x}$ is a continuous
random variable which requires infinitely many bits to be specified
with unlimited precision. If there is an information bottleneck,
however, only a limited amount of bits can be transmitted. Both, the
multi-information and the average log-loss do not take into account
the problem what information should be encoded and what information
can be discarded.  Therefore, a representation optimized only for
redundancy reduction can in fact perform quite poorly with respect to
the goal of preserving the {\it relevant} information during the
transmission \citep{bethge:2006,goyal:2001,gray:1990}. In order to
compare two different representations with respect to this more
complete notion of coding efficiency, it is necessary to have a
measure for the quality with which a signal can be reconstructed from
the information that is preserved by the representation.  This
objective is in fact very much related to the task of image
compression.

Clearly, we expect that the criteria for judging image compression
algorithms may not provide a good proxy to an accurate judgement
of what information is considered relevant in a biological vision system.
In particular, the existence of selective attention suggests that different
aspects of image information are transmitted at different times depending
on the behavioral goals and circumstances \citep{li:2006}. That is,
a biological organism can change the relevance criteria dynamically
on demand while for still image compression algorithms it is rather
necessary that this assessment is made once and forever in a fixed
and static fashion.  

The issue of selective attention is outside the scope of this paper.
Striving for simplicity instead, we will use the mean squared
reconstruction error for the pixel intensities. This is the measure of choice
for high-rate still image compression \cite{wang:2002}.  In particular,
it is common to report on the performance of a code by determining
its rate--distortion curve which specifies the required information rate for a
given reconstruction error (and vice versa) \cite{gray:1990}.  Consequently,
we will ask for a given information rate, how do the image representations
compare with respect to the reconstruction error.  As result, we will obtain a
so-called rate--distortion curve which displays the average
reconstruction error as a function of the information rate or vice
versa. The second method used in \cite{lewicki:1999,lewicki:2000} is
an estimate of a single point on this curve for a particular fixed
value of the reconstruction error.

The estimation of the rate--distortion curve is clearly the most
difficult task among the three criteria. We adopt the framework of
transform coding\citep{goyal:2001} extensively used in still image compression,
which divides the encoding task
into two steps: First, the image patches $\mathbf x$ are linearly
transformed into $\mathbf y = W\mathbf x$. Then the coefficients $y_j$
are quantized independently of each other.  Using this framework, we
can ask whether the use of an ICA image transformation leads to a
smaller reconstruction error after coefficient quantization than PCA
or any other transform.

As for quantizing the coefficients, we resort to the framework of
variable rate entropy coding \citep{gray:1998}. In particular, we
apply uniform quantization, which is close to optimal for high-rate
compression \citep{gish:1968,goyal:2001}. For uniform quantization,
it is only required to specify the bin width of the coefficients. There is also
the possibility to use a different number of quantization levels for
the different coefficients. The question of how to set these numbers
is known as the `bit allocation problem' because the amount of bits
needed to encode one coefficient will depend monotonically on the
number of quantization levels. The number of quantization levels can
be adjusted in two different but equivalent ways: One possibility is
to use a different bin width for each individual
coefficient. Alternatively, it is also possible to use the same bin
width for all coefficients and multiply all coefficients with an
appropriate scale factor before quantization. The larger the variance
of an individual coefficient, the more bits will be allocated to
represent it.

Here, we will employ the latter approach, for which the bit allocation
problem becomes an inherent part of the transformation: Any bit
allocation scheme can be obtained via post-multiplication with a
diagonal matrix. Thus, in contrast to the objective function of ICA,
the rate--distortion criterion is not invariant against
post-multiplication with a diagonal matrix. For ICA and PCA, we will
determine the rate--distortion curve for both, normalized output
variances (``white ICA'' and ``white PCA'') and normalized basis
functions (``normalized ICA'' and ``orthonormal PCA''), respectively.


\subsection*{Decorrelation transforms}
\label{sec:DecorrelationTransforms}
The particular shape of the ICA basis functions is obtained by
minimization of the multi-information over all invertible linear
transforms $\mathbf{y}=W\mathbf{x}$. In contrast, the removal of
second-order correlations alone generally does not yield localized,
oriented, and bandpass image basis functions. ICA additionally removes
higher-order correlations which are generated by linear mixing.  In
order to assess the importance of this type of higher-order
correlations for redundancy reduction and coding efficiency we will
compare ICA to other decorrelating image bases.

Let $C = \expc{\mathbf{xx}^\top}$ be the covariance matrix of the data
and $C = UDU^\top$ its eigen-decomposition.  Then, any linear
second-order decorrelation transform can be written as
\begin{equation}
  W \,=\, D_2 \cdot V \cdot D^{-1/2} \cdot U^\top
\end{equation}
where $D$ and $U$ are defined as above, $V$ is an arbitrary orthogonal
matrix and $D_2$ is an arbitrary diagonal matrix.  It is easily
verified that $Y = WX$ has diagonal covariance for all choices of $V$
and $D_2$, i.e. all second-order correlations vanish.  This means that
any particular choice of $V$ and $D_2$ determines a specific
decorrelation transform. Based on this observation we introduce a
number of linear transformations for later reference. All matrices are
square and are chosen to be of determinant $\lambda^m$, where $m$ is
the number of columns (or rows) of $W$ (i.e.\ $\lambda=\sqrt[m]{\prod
  \lambda_i}$ is the geometrical mean of the eigenvalues $\lambda_i,
i=1, \dots, m$).
\begin{description}
\item[Orthogonal principal component analysis (oPCA)]~\\
  If the variances of the principle components (i.e. the diagonal
  elements of $D$) are all different, PCA is the only
  metric-preserving decorrelation transform and is heavily used in
  digital image coding. It corresponds to choosing $V = I_m$ as the
  identity matrix and $D_2 = \lambda D^{1/2}$, such that
  $W_\text{oPCA}=\lambda U^\top$.
  
\item[White principal component analysis (wPCA)]~\\
Equalizing the output variances in the PCA representation sets the stage
for the derivation of further decorrelation transforms different from PCA.
In order to assess the effect of equalization for coding efficiency, we also
include this ``white PCA'' representation into our analysis:
  Choose $V = I_m$ as for orthonormal PCA and then set $D_2= \mu \,
  I_m$ with $\mu=\lambda \sqrt[m]{\det(D^{1/2})}$ such that
  $W_\text{wPCA}= \mu\,\, D^{-1/2}U^\top$.
  
\item[Symmetric whitening (SYM)]~\\
  Among the non-orthogonal decorrelation transforms, symmetric
  whitening stays as close to the input representation as possible (in
  Frobenius norm) \cite{fan:1955}.  In terms of early vision this may
  be seen as an implementation of a wiring length minimization
  principle. Remarkably, the basis functions of symmetric whitening
  resemble the center-surround shape of retinal ganglion cell
  receptive fields when applied to the pixel representation of natural
  images \cite{atick:1992a}. The symmetric whitening transform is
  obtained by setting $V=U$ and $D_2= \mu \, I_m$ such that
  $W_\text{SYM}= \mu \,\, U D^{-1/2}U^\top$.

\item[Random whitening (RND)]~\\
  As a baseline which neither exploits a special structure with
  respect to the input representation nor makes use of higher-order correlations
  we also consider a completely random transformation.
  To obtain a random orthogonal matrix we first draw a random matrix
  $G$ from a Gaussian matrix-variate distribution and then we set
  $V_\text{RND} =(GG^\top)^{-1/2} G$. With $D_2= \mu \, I_m$ we obtain
  $W_\text{RND} = \mu \,\, V_\text{RND} D^{-1/2}U^\top$.

\item[White independent component analysis (wICA)]~\\
  Finally, ICA is the transformation which has been suggested to explain
  the orientation selectivity of V1 simple cells \cite{olshausen:1996,bell:1997}.
  Set $V=V_\text{ICA}$ for which the multi-information $I[Y]$ takes a
  minimum. With $D_2= \mu \, I_m$ we obtain $W_\text{wICA} = \mu \,\,
  V_\text{ICA} D^{-1/2}U^\top$.

\item[Normalized independent component analysis (nICA)]~\\
  Normalized independent component analysis (nICA) differs from white
  ICA ($W_\text{wICA}$) only by a different choice of the second
  diagonal matrix $D_2$. Instead of having equal variance in each
  coefficient, we now choose $D_2$ such that the corresponding basis
  vector of each coefficient has the same length in pixel space. It is
  easy to see that our first two criteria, the multi-information and
  the negative log-likelihood, are invariant under changes in $D_2$.
  It makes a difference for the rate--distortion curves as in our
  setup the variance (or, more precisely, the standard deviation)
  determines the bit allocation.  Practically, $W_\text{nICA}$ can be
  determined by using $W_\text{wICA}$ as follows: First, we compute
  the matrix inverse $A:=W_\text{wICA}^{-1}$ and determine the
  Euclidean norm $a_1, \dots, a_m$ of the column vectors of $A$. With
  $D_a= \mbox{diag}(a_1, \dots,a_m)$, we then obtain $W_\text{nICA} =
  \frac{1}{\sqrt[m]{\det(D_a)}} D_a W_\text{wICA}$.

\end{description}

\subsection*{ICA algorithm}
\label{sec:ICAalgorithm}
If the true joint probability distribution is known, the minimization
of the multi-information over all linear transformations can be
formulated without any assumptions about the shape of the
distribution. In practice, the multi-information has to be estimated
from a finite amount of data which requires to make assumptions about
the underlying density.

There are many different ICA algorithms which differ in the
assumptions made and also in the optimization technique employed.  The
choice of the particular ICA algorithm used here was guided by a set
of requirements that arise from the specific problem setting.
Although a wide variety of ICA algorithms has been published, none of
them fits exactly all of our requirements.

We would like to use an ICA algorithm, which gives the ICA image basis
the best chance for the comparison with other image representations.
For the comparison of the multi-information reduction, we are using
the OPT estimator introduced in \citep{bethge:2006} which has been
found to give the most reliable results. This estimator employs a
parametric estimate of the coefficient distributions based on the
exponential power family which is known to provide an excellent fit to
the coefficient distributions of natural images
\citep{srivastava:2003,bethge:2006}. Our ICA algorithm should make the
same assumptions about the data as we make for the final comparison of
the multi-information reduction.  Therefore, we are also using the
exponential power family model for the marginal densities during the
minimization of the multi-information. In addition, we want to have an
ICA basis which is indistinguishable from the other image
representations with respect to the second-order statistics.
Therefore, we are using a pre-whitened ICA algorithm, whose search
space is restricted to the subgroup of orthogonal matrices $SO(n)$.
One of the most efficient ICA methods in the public domain specialized
to pre-whitened ICA is FastICA \citep{hyvarinen_book:2001}.  We
use this fixed-point algorithm as an initialization.  Subsequently,
the solution is further refined by performing a gradient ascent over
the manifold of orthogonal matrices on the likelihood of the data, when
each marginal is modelled by a the exponential power distribution as
in the case of the OPT estimator.

In order to optimize the objective function over the subspace of
orthogonal matrices, we adapted the algorithms for Stiefel manifolds
proposed by Edelman {\it et al.} \citep{EdeAriSmi1999} to the simpler
case of orthogonal groups and combined it with the line-search routine
\texttt{dbrent} from \citep{NumRecC1992} to achieve a rather
straightforward gradient descent algorithm.  For the initialization
with FastICA, we use the Gaussian non-linearity, the symmetric
approach and a tolerance level of $10^{-5}$.

\subsection*{Spherically symmetric model}

A well known result by Maxwell \citep{maxwell:1855} states that the
only factorial distribution invariant against arbitrary orthogonal
transformations is the isotropic Gaussian distribution. Natural images
exhibit marginals which are significantly more peaked than
Gaussian. Nevertheless, their distribution does share the spherical
symmetry with the Gaussian as already found by \cite{zetzsche:1999}
for gabor filter pairs and lately exploited by \cite{lyu:2008} for
nonlinear image representations. Therefore, it makes sense to compare
the performance of the ICA model with a spherically symmetric model of
the whitened data $\MB{y}_\text{w}=W_\text{RND}\MB{x}$. Note that any
spherically symmetric model is still invariant under orthogonal
transformations while only the Gaussian additionally exhibits marginal
independence.

While the radial distribution of a Gaussian (i.e. the distribution
over the lengths of the random vectors) is a $\chi$ distribution,
whose shape and scale parameter is determined by the number of
dimensions and the variance, respectively, the spherical symmetric
model may be seen as a generalization of the Gaussian, for which the
radial distribution $p(r)$ with $r:=||\mathbf{y}||_2$ can be of
arbitrary shape. The density of the spherically symmetric distribution
(SSD) is defined as $p_{\mathbf y}(\MB{y})=p_r(r)/S_n(r)$, where
$S_n(r)=r^{n-1}2\pi^{n/2}/\Gamma(n/2)$ is the surface area of a sphere
in $\mathbb R^n$ with radius $r$. For simplicity we will model the
radial distribution with a member of the Gamma family
\begin{equation}
  \label{eq:SSD}
  p(r)
  \,=\,
    \frac{r^{u-1} \exp\left(-\frac{r}{s}\right)}{s^u \Gamma(u)} \quad, \, r \ge 0
\end{equation}
with shape parameter $u$ and scale parameter $s$, which can be easily
matched to the mean and variance of the empirical distribution via $s=
\widehat{\text{Var}}[r]/\expe{r}$ and $u=\expe{r}^2/
\widehat{\text{Var}}[r]$.

\subsection*{Dataset}
\label{sec:Dataset}

The difference in the performance between ICA and other linear
transformations clearly depends on the data. For {\it gray-scale}
images we observed in our previous study \citep{bethge:2006} that the
difference in the multi-information between ICA and any other
decorrelation transform is consistently smaller than 5\%. In
particular, we controlled for the use of different pictures and for
the effect of different pre-processing steps.

Here, we resort to the dataset used in a previous study
\citep{wachtler:2001,LeeWacSej2002}, which among all previous studies
reported the largest advantage of ICA compared to PCA. This {\it
  color} image dataset is based on the Bristol Hyperspectral Images
Database \citep{brelstaff:1995} that contains multi-spectral
recordings of natural scenes taken in the surroundings of Bristol, UK
and in the greenhouses of Bristol Botanical Gardens. The authors of
\citep{LeeWacSej2002} kindly provided to us a pre-processed version
of the image data where spectral radiance vectors were already
converted into LMS values.  During subsequent
processing the reflectance standard was cut out and images were
converted to log intensities (cf.~\citep{LeeWacSej2002}).

All images come at a resolution of $256\times256$ pixels.  From each
image circa 5000 patches of size $7\times7$ pixels were drawn at
random locations (circa 40000 patches in total).  For chromatic images
with three color channels (LMS) each patch is reshaped as a
$7\times7\times3=147$-dimensional vector.  To estimate the
contribution of color information, a comparison with monochromatic
images was performed where gray-value intensities were computed as $I
= \log(\frac{1}{3}(L+M+S))$ and exactly the same patches were used for
analysis.  In the latter case, the dimensionality of a data sample is
thus reduced to 49 dimensions. Our motivation to chose $7\times7$
patches is to ensure maximal comparability to the study of
\citep{LeeWacSej2002}. We carried out the same analysis for patch
sizes of $15\times 15$ as well. For the sake of clarity, only the
number for the $7\times 7$ patches are reported in this paper. The
results for $15\times 15$ can be found in the additional material.
All experiments are carried out over ten different training and test
sets sampled independently from the original images. 

Since the statistics of the constant illumation part of image patches,
i.e. the DC component, differs significantly from the statistics of
luminance variations, we removed the DC component from the patches
before further transforming them. In order to not affect the entropy
of the data, we used an orthogonal transformation. The projector
$P_{remDC}$ is computed such that the first (for each color channel)
component of $P_{remDC}\mathbf{x}$ corresponds to the DC component(s)
of that patch. One such a possible choice is the matrix $$
P=\begin{pmatrix}1 & 0 & 0 & \cdots\\
  1 & 1 & 0 & \cdots\\
  1 & 0 & \ddots & \cdots\\
  \vdots & & & 1\end{pmatrix}^\top $$ However, this is not an
orthogonal transformation. Therefore, we decompose $P$ into $P=QR$
where $R$ is upper triangular and $Q$ is an orthogonal transform.
Since $P=QR$, the first column of $Q$ must be a multiple of the vector
with all coefficients equal to one (due to the upper triangluarity of
$R$).  Therefore, the first component of $Q^{\top}\mathbf{x}$ is a
multiple of the DC component. Since $Q$ is an orthonomal transform,
using all but the first row of $Q^{\top}$ for $P_{\text{remDC}}$
projects out the DC component. In the case of color images
$P_{\text{remDC}}$ becomes a block-diagonal matrix with $Q^{\top}$ as
diagonal elements for each channel.

By removing the DC component in that manner, all linear
transformations are applied in $n-1$ dimensions, if $n$ denotes the
number of pixels in the original image patch. In this case the
marginal entropy of the DC-components has to be included in the
computation of the multi-information in order to ensure a valid
comparison with the original pixel basis. We use the same estimators
as in \citep{bethge:2006} to estimate the marginal entropy of
DC-component.

\section*{Results}
\label{Sec:results}

\subsection*{Filter Shapes}

\begin{figure}[H]
  \begin{center}
    \includegraphics{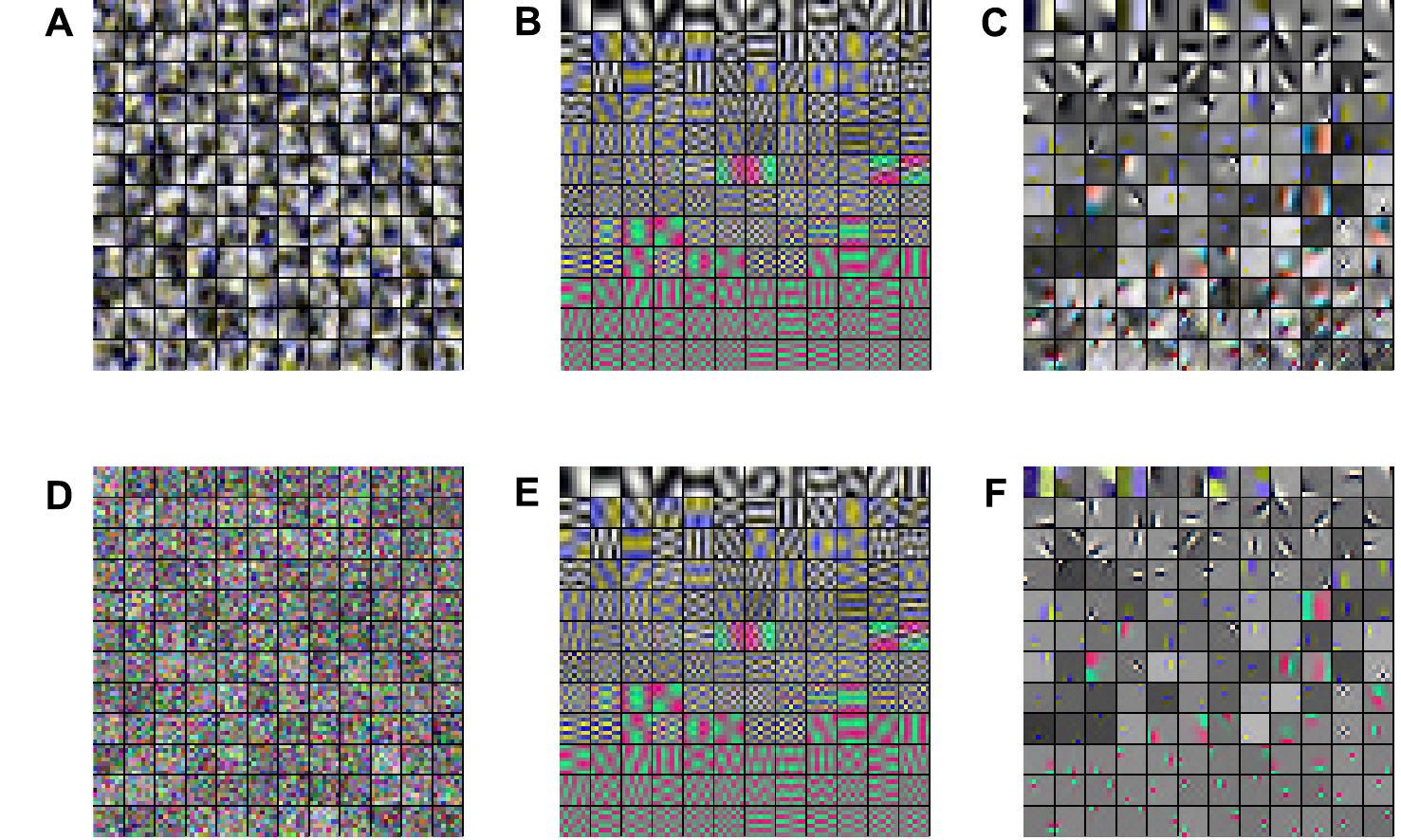}
  \end{center}
  \caption{{\bf Examples for Receptive Fields of Various Image
      Transforms } \hspace{0.3cm} Basis functions of a random
    decorrelation transform ({\bf\textsf{RND}}), principal component
    analysis ({\bf\textsf{PCA}}) and independent component analysis
    ({\bf\textsf{ICA}}) in pixel space
    ({\bf\textsf{A}}-{\bf\textsf{C}}) and whitened space
    ({\bf\textsf{E}}-{\bf\textsf{F}}). The image representation in
    whitened space is obtained by left multiplication with the matrix
    square root of the inverse covariance matrix $C^{-1/2}$.  This
    figure can only give a rough idea of the shape of the basis
    functions. For a detailed inspection of the basis functions we
    refer the reader to our web page
    \url{http://www.kyb.mpg.de/bethge/code/QICA/} where we provide all
    the data and code used in this paper.}
  \label{fig:BasisFunctions}
\end{figure}

As in previous studies \citep{olshausen:1996,bell:1997} the filters
derived with ICA exhibited orientation selective tuning properties
similar to those observed for V1 simple cells (see
Figure~\ref{fig:BasisFunctions}). For illustration, we also show the
basis functions learned with PCA and RND in
Figure~\ref{fig:BasisFunctions}. The basis functions $A$ are obtained by
inverting the filter matrix $W$ (including the DC component).  The
result is displayed in the upper panel
(Figure~\ref{fig:BasisFunctions} {\bf{\textsf{A-C}}}). Following
common practice, we also visualize the basis functions after symmetric
whitening (Figure~\ref{fig:BasisFunctions} {\bf{\textsf{D-F}}}). 

The basis functions of both PCA and ICA exhibit color opponent coding
but the basis functions of ICA are additionally localized and orientation selective.
The basis functions of the random decorrelation transform does not
exhibit any regular structure besides the fact that they are bandpass.
The following quantitative comparisons will show, however, that the
distinct shape of the ICA basis functions does not yield a clear advantage
for redundancy reduction and coding efficiency.

\subsection*{Multi-information}

\begin{figure}[H]
  \begin{center}
  \includegraphics{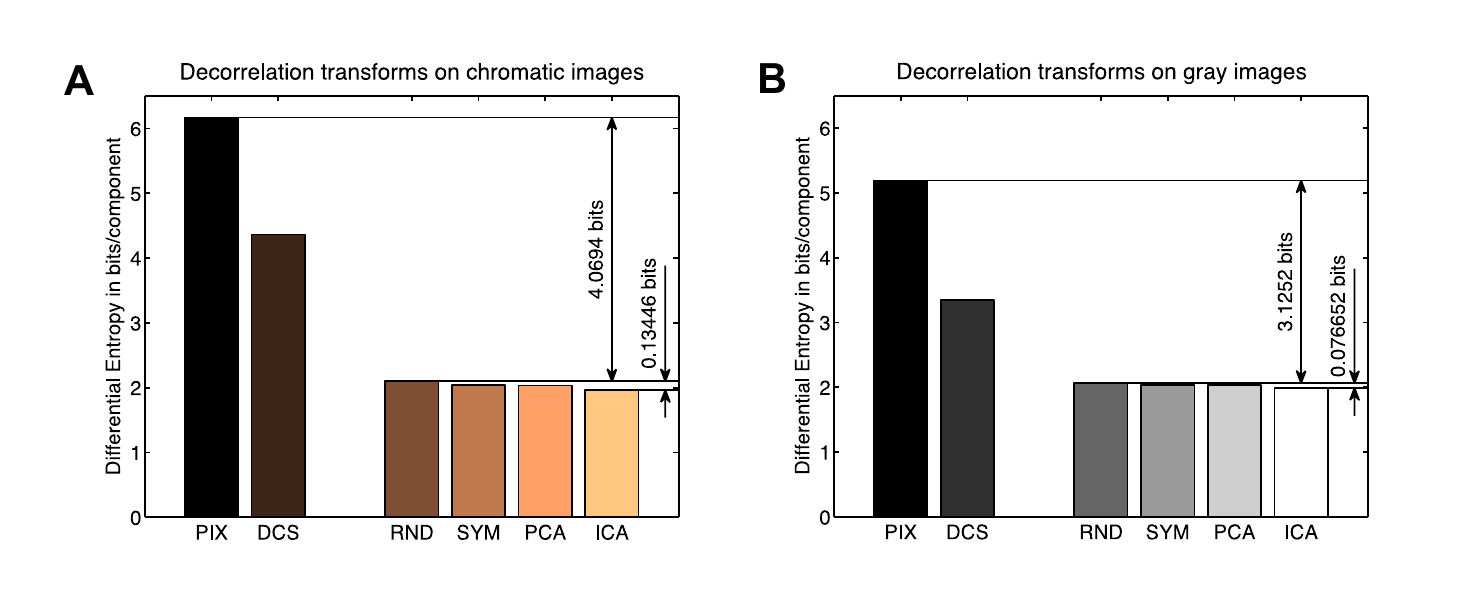}
  \end{center}
  \caption{{\bf Multi-Information Reduction per
      Dimension}\hspace{0.3cm} Average differential entropy $\langle h
    \rangle$ for the pixel basis (PIX), after separation of the DC
    component (DCS), and after application of the different
    decorrelation transforms. The difference between PIX and RND
    corresponds to the redundancy reduction that is achieved with
    a random second-order decorrelation transform. The small
    difference between RND and ICA is the maximal amount of
    higher-order redundancy reduction that can be achieved by ICA.
    Diagram ({\bf\textsf{A}}) shows the results for
    chromatic images and diagram ({\bf\textsf{B}}) for gray value
    images. For both types of images, only a marginal amount can
    be accounted to the reduction of higher order dependencies.}
  \label{fig:MultiInfoBarplot}
\end{figure}

The multi-information is the true objective function that is minimized by
ICA over all possible linear decorrelation transforms.
Figure~\ref{fig:MultiInfoBarplot} shows the reduction in
multi-information achieved with different decorrelation transforms
including ICA for chromatic and gray value images, respectively. For
each representation, the results are reported in bits per component,
i.e. as marginal entropies averaged over all dimensions:
\begin{equation}
\langle h \rangle = \frac{1}{n} \sum_{k=1}^n h[p_k(y_k)]
\end{equation}

Table~\ref{tab:MultiInfo} shows the corresponding values for the
transformations RND, SYM, PCA and ICA. For both chromatic images and
gray-value intensities, the lowest and highest reduction is achieved
with RND or ICA, respectively. However, the additional gain in the
multi-information reduction achieved with ICA on top of RND constitutes only
3.20\% for chromatic images and 2.39\% for achromatic in comparison with the
total reduction relative to the pixel basis (PIX). This means that
only a small fraction of redundancy reduction can actually be accounted
to the removal of higher-order redundancies with ICA.

\begin{table}[H]
  \small
  \centering
  \renewcommand{\arraystretch}{1.8}

\begin{tabular}{|c|c|c||c|c|c|}
\hline
\multicolumn{3}{|c||}{\textbf{Absolute Difference}} & \multicolumn{3}{c|}{\textbf{Relative Difference}}\tabularnewline
\hline
\hline
 & Color & Gray &  & Color & Gray\tabularnewline
\hline
RND-PIX & -4.0694 $\pm$ 0.0043 & -3.1252 $\pm$ 0.0043 &  &  & \tabularnewline
\hline
SYM-RND & -0.0593 $\pm$ 0.0004 & -0.0259 $\pm$ 0.0006 & $\frac{\mbox{SYM-RND}}{\mbox{SYM-PIX}}$ & 1.44 $\pm$ 0.01 & 0.82 $\pm$ 0.02\tabularnewline
\hline
PCA-RND & -0.0627 $\pm$ 0.0008 & -0.0353 $\pm$ 0.0011 & $\frac{\mbox{PCA-RND}}{\mbox{PCA-PIX}}$ & 1.52 $\pm$ 0.02  & 1.12 $\pm$ 0.03\tabularnewline
\hline
ICA-RND & -0.1345 $\pm$ 0.0008 & -0.0767 $\pm$ 0.0008 & $\frac{\mbox{ICA-RND}}{\mbox{ICA-PIX}}$ & 3.20 $\pm$ 0.02  & 2.39 $\pm$ 0.02\tabularnewline
\hline
\end{tabular}

  \caption{{\bf Comparision of the Multi-Information Reduction for
      Chromatic and Achromatic Images}\hspace{0.3cm} Differences in the multi-information reduction between various
    decorrelation transforms (SYM, PCA, ICA) relative to a random decorrelation
    transform (RND) compared to the multi-information reduction achieved with
    the random decorrelation transform relative to the original
    pixel basis (RND-PIX). The {\it absolute} multi-information reduction is given in
    bits/component on the left hand side. How much more the special decorrelation
    transforms SYM, PCA and ICA can reduce the multi-information {\it relative} to the
    random (RND) one is given in percent on the right hand side.}
  \label{tab:MultiInfo}
\end{table}

\begin{figure}[H]
  \begin{center}
    \includegraphics{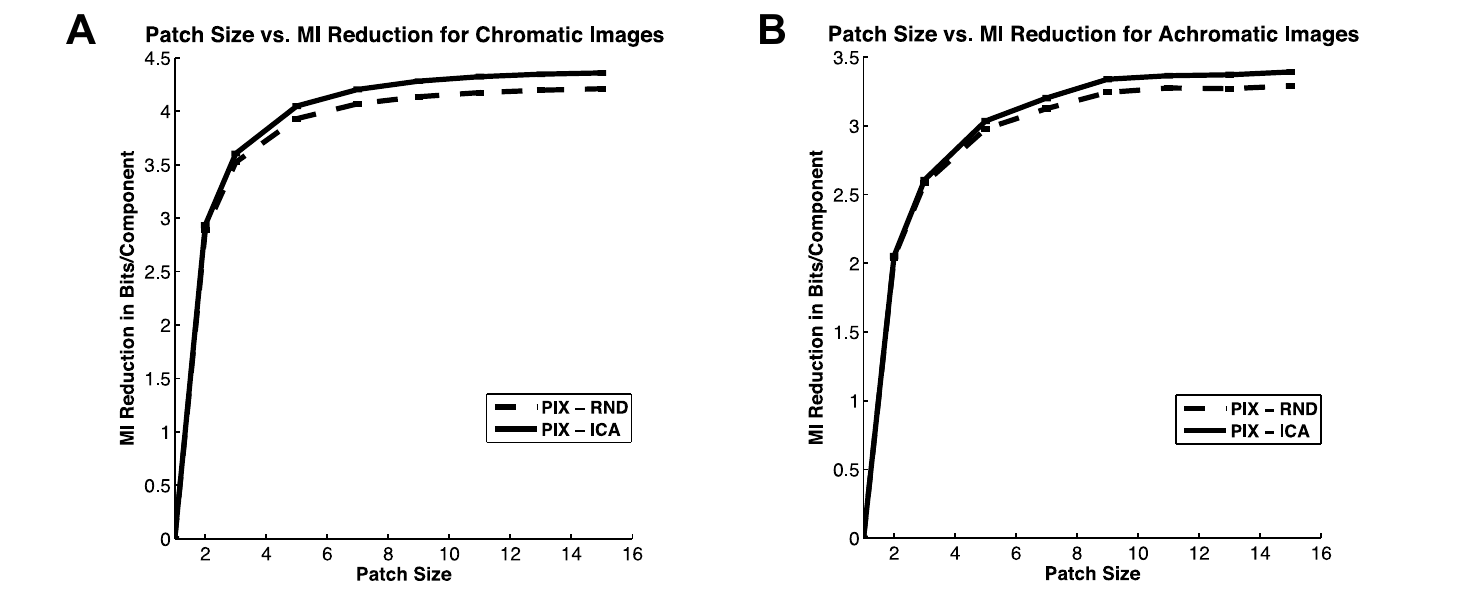}
  \end{center}
  \caption{{\bf Patch Size vs. Redundancy Reduction} \hspace{0.3cm}
    Redundancy reduction as a function of the patch size. The graph
    shows the multi-information reduction with respect to the pixel
    representation (PIX) achieved by the transformations RND and ICA.
    The achievable reduction quickly saturates with increasing patch
    size such that its value for $7\times 7$ image patches is already
    at about 90\% of its asymptote. In particular, the small advantage
    of ICA over other transformation does not only occur for small
    patch sizes for which spatial localization is imposed on the basis
    functions for all transformations due to the patch boundaries.}
  \label{fig:ptchSizeVsMIRed}
\end{figure}

One may argue that the relatively small patch size of $7\times 7$
pixel may be responsible for the small advantage of ICA as all
decorrelation functions already getting the benefit of
localization. 
In order to address the question how the patch size affects the linear
redundancy reduction, we repeated the same analysis on a whole range
of different patch sizes. Figure~\ref{fig:ptchSizeVsMIRed} shows the
multi-information reduction with respect to the pixel representation
(PIX) achieved by the transformations RND and ICA. The achievable
reduction quickly saturates with increasing patch size such that its
value for $7\times 7$ image patches is already at about 90\% of its
asymptote. In particular, one can see that the relative advantage of ICA
over other transformations is still small ($\sim$ 3\%) also for large patch sizes.
All Tables and Figures for patch size $15\times 15$ can be found in
the additional material.

\subsection*{Average log-loss}

Since redundancy reduction can also be interpreted as a special form
of density estimation we also look at the average log-loss which quantifies
how well the underlying density model of the different transformations
is matched to the statistics of the data. Table~\ref{tab:AverageLogLoss}
shows the average log-loss (ALL) and the differential log-likelihood (DLL)
in bits per component. 
For the average log-loss, ICA achieved an ALL of $1.78$ bits per
component for chromatic images and $1.85$ bits per component for
achromatic images.  Compared to the ALL in the RND representation of
$1.94$ bits and $1.94$ bits, respectively, the gain achieved by ICA is again
small.  Additionally, the ALL values were very close to the
differential entropies, resulting in small DLL values. This confirms
that the exponential power distribution fits the shape of the
individual marginal coefficient distributions well.  Therefore, we can
safely conclude that the advantage of ICA is small not only in terms
of redundancy reduction as measured by the multi-information, but also
in the sense of density estimation.

\begin{table}[H]
  \centering
  \renewcommand{\arraystretch}{1.1}

\begin{tabular}{|c|c|c|c|c|}
\hline
 & \multicolumn{2}{c|}{\textbf{Color}} & \multicolumn{2}{c|}{\textbf{Gray}}\tabularnewline
\hline
\hline
\textsf{\textbf{A}} & \multicolumn{2}{c|}{ALL} & \multicolumn{2}{c|}{ALL}\tabularnewline
\cline{2-3} \cline{4-5}
\hline
RND & \multicolumn{2}{c|}{1.9486 $\pm$ 0.0035} & \multicolumn{2}{c|}{1.9414 $\pm$ 0.0044}\tabularnewline
\hline
SYM-RND & \multicolumn{2}{c|}{-0.0881 $\pm$ 0.0004} & \multicolumn{2}{c|}{-0.0402 $\pm$ 0.0005}\tabularnewline
\hline
PCA-RND & \multicolumn{2}{c|}{-0.0751 $\pm$ 0.0009} & \multicolumn{2}{c|}{-0.0391 $\pm$ 0.0011}\tabularnewline
\hline
ICA-RND & \multicolumn{2}{c|}{-0.1637 $\pm$ 0.0007} & \multicolumn{2}{c|}{-0.0880 $\pm$ 0.0007}\tabularnewline
\hline
SSD-RND & \multicolumn{2}{c|}{-0.2761 $\pm$ 0.0025} & \multicolumn{2}{c|}{-0.2868 $\pm$ 0.0032}\tabularnewline
\cline{2-3}
\hline
\textsf{\textbf{B}} & DLL & $\langle\alpha\rangle$ & DLL & $\langle\alpha\rangle$\tabularnewline
\hline
RND & -0.0113 $\pm$ 0.0007 & 1.0413 $\pm$ 0.0026 & -0.0057 $\pm$ 0.0006 & 1.0132 $\pm$ 0.0046\tabularnewline
\hline
SYM & -0.0388 $\pm$ 0.0009 & 0.8961 $\pm$ 0.0021 & -0.0195 $\pm$ 0.0009 & 0.9486 $\pm$ 0.0040\tabularnewline
\hline
PCA & -0.0224 $\pm$ 0.0007 & 0.9145 $\pm$ 0.0024 & -0.0087 $\pm$ 0.0007 & 0.9425 $\pm$ 0.0025\tabularnewline
\hline
ICA & -0.0378 $\pm$ 0.0009 & 0.7687 $\pm$ 0.0017 & -0.0154 $\pm$ 0.0011 & 0.8434 $\pm$ 0.0025\tabularnewline
\hline
\end{tabular}

\caption{{\bf Comparision of the Average Log-Loss (ALL) and the
    Differential Log-Likelihood (DLL) Chromatic and Achromatic Images}
  \hspace{0.3cm} {\bf\textsf{A.}} The first row shows the average log-loss (ALL, in 
  bits/component) of the density model determined by the linear
  transformation RND. The value was obtained by averaging over 10
  separately sampled training and test sets of size 40.000 and
  50.000, respectively. The following rows shows the difference
  of the ALL of the models SYM, PCA, ICA and the spherically
  symmetric density (SSD) to the ALL determined by linear
  transformation RND. The large value for RND$-$ICA fundamentally
  contradicts the assumptions underlying the ICA model.
  {\bf\textsf{B.}} The small DLL values suggest, that
  the exponential power distribution fits the shape of the individual
  coefficient distributions well. In addition, we also report the
  average exponent $\langle \alpha \rangle$ of the exponential power
  family fit to the individual coefficient distributions ($\alpha =
  1$ corresponds to a Laplacian shape).}  
  \label{tab:AverageLogLoss}
\end{table}

\paragraph{Comparison to a spherical symmetric model}
The fact that ICA fits the distribution of natural images only
marginally better than a random decorrelation transform implies that
the generative model underlying ICA does not apply to natural images.
In order to assess the importance of the actual filter shape, we
fitted a spherically symmetric model to the filter responses. The
likelihood of such a model is invariant under post-multiplication of
an orthogonal matrix, i.e. the actual shape of the filter. Therefore,
a good fit of such a model provides strong evidence against a
critical role of certain filter shapes.

As shown in Table~\ref{tab:AverageLogLoss}, the ALL of the SSD model
is $1.67$ bits per component for chromatic images and $1.65$ bits per
component for achromatic images. This is significantly smaller than the ALL of
ICA indicating that it fits the distribution of natural images much
better than ICA does.  This result is particularly striking if one compares the
number of parameters fitted in the ICA model compared to the SSD case:
After whitening, the optimization in ICA is done over the manifold of
orthogonal matrices which has $m(m-1)/2$ free parameters (where $m$ denotes
the number of dimensions without the DC components). The additional optimization of
the shape parameters for the exponential power family fitted to each
individual component adds another $m$ parameters. For the case of
$7\times 7$ color image patches we thus have $\frac{144\cdot
  145}{2}=10440$ parameters. In stark contrast, there are only two
free parameters in the SSD model with a radial Gamma distribution, the
shape parameter $u$ and the scale parameter $s$. Nevertheless, for
chromatic images the gain of the SSD model relative to random
whitening is almost twice as large as that of ICA and even three and a
half times as large for achromatic images.

Since the SSD model is completely independent of the choice of the orthogonal
transformation after whitening, its superior performance compared with ICA provides
a very strong argument against the hypothesis that orientation selectivity plays
a critical role for redundancy reduction. In addition, it is also corroborates earlier
arguments that has been given to show that the statistics of natural images
does not conform to the generative model underlying ICA \cite{zetzsche:1999, baddeley:1996}.

\begin{figure}[H]
  \begin{center}
  \includegraphics{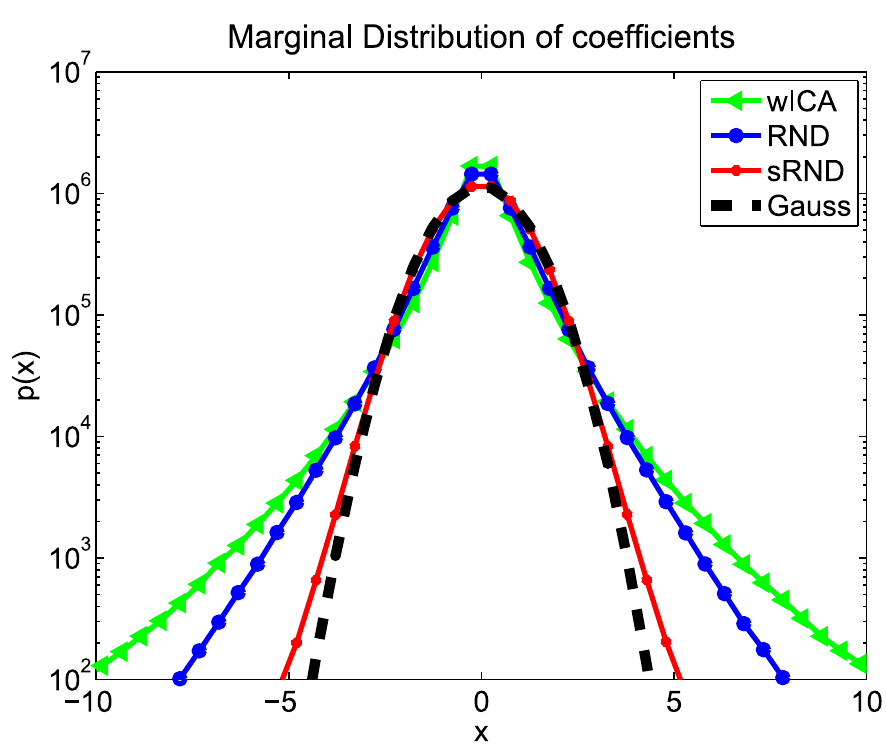}
  \end{center}
  \caption{\label{Fig:sRND}{\bf The Distribution of Natural Images
      does not Conform with the Generative Model of ICA}\hspace{0.3cm}
    { In order to test for statistical dependencies among the
      coefficients $Y_\text{wICA}$ of whithened ICA for single data
      samples, the coefficients were shuffled among the data points
      along each dimension.  Subsequently, we transform the resulting
      data matrix $Y_\text{sICA}$ into $Y_\text{sRND} =
      W_{RND}W_{wICA}^{-1} Y_\text{sICA}$. This corresponds to a
      change of basis from the ICA to the random decorrelation basis
      (RND).  The plot shows the log-histogram over the coefficients
      over all dimensions.  If the assumptions underlying ICA were
      correct, there would be no difference between the histogram of
      $Y_\text{sRND}$ and $Y_\text{RND}$.}}
\end{figure}

Besides the better fit of the data by the SSD model, there is also a
more direct way of demonstrating the dependencies of the ICA coefficients: If
$Y_\text{wICA}=(\mathbf{y}_1, \dots, \mathbf{y}_N)$ is data in the
wICA representation, then the independence assumption of ICA can be
simulated by applying independent random permutations to the rows of
$Y_\text{wICA}$. Certainly, such a shuffling procedure does not alter
the histograms of the individual coefficients but it is suited to
destroy potential statistical dependencies among the coefficients.
Subsequently, we can transform the shuffled data $Y_\text{sICA}$ back
to the RND basis $Y_\text{sRND}=W_\text{RND}W_\text{wICA}^{-1}
Y_\text{sICA}$.  If the ICA coefficients were independent, the
shuffling procedure would not alter the joint statistics, and hence,
one should find no difference in the multi-information between
$Y_\text{sRND}$ and $Y_\text{RND}$.  But infact, we observe a large
discrepancy between the two (Figure~\ref{Fig:sRND}). The
distributions of the sRND coefficients were very close to Gaussians
and the average marginal entropy of sRND yielded $\langle
h_\text{sRND} - h_\text{Gauss} \rangle \approx -0.001$~bits in
contrast to $\langle h_\text{RND} - h_\text{Gauss} \rangle \approx
-0.1$~bits. In other words, the finding that for natural images the
marginals of a random decorrelation transform have Laplacian shape
($\alpha \approx 1$) stands in clear contradiction to the generative
model underlying ICA. If the ICA model was valid, one would expect
that the sum over the ICA coefficients would yield Gaussian marginals
due to the central limit theorem. In conclusion, we have very strong
evidence that the ICA coefficients are not independent in case of
natural images.

\subsection*{Rate-distortion curves}
\label{sec:RateDistortion}

There are different ways to account for the information bottleneck that is
imposed by neural noise and  firing rate limitations. The advantage with respect to
a plain information maximization criterion can equivalently be measured by the
multi-information criterion considered above. In order to additionally account for the question
which representation encodes most efficiently the {\it relevant} image information,
we also present rate distortion curves which show the minimal reconstruction
error as a function of the information rate.

\begin{figure}[H]
   \begin{center}
     \includegraphics{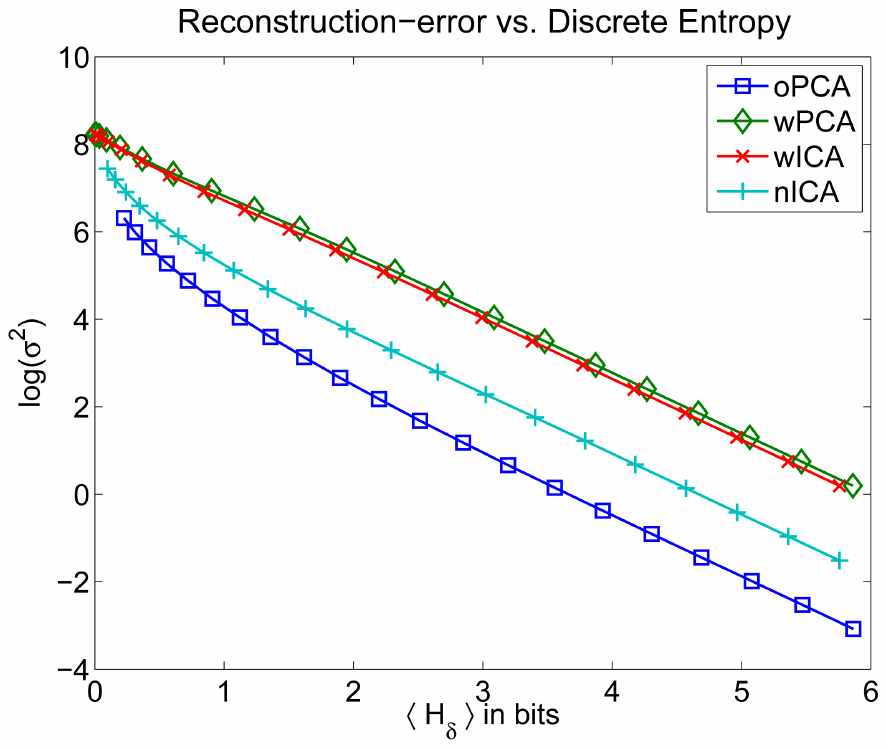}
     \end{center}
     \caption{{\bf Rate-distortion Curves } \hspace{0.3cm}
       Rate-distortion curve for PCA and ICA when equalizing the
       output variances (wPCA and wICA) and when equalizing the norm
       of the corresponding image bases in pixel space (oPCA and
       nICA). The plot shows the discrete entropy $H_\delta$ in bits
       (averaged over all dimensions) against the log of the squared
       reconstruction error $\sigma^2$. oPCA outperforms all other
       transforms in terms of coding efficiency. wPCA in turn
       performes worst and remarkably similar to wICA. Since wPCA and
       wICA differ only by an orthogonal transformation, both
       representations are bound to the same metric.  oPCA is the only
       transformation which has the same metric as the pixel
       representation according to which the reconstruction error is
       determined. By normalizing the length of the ICA basis vectors
       in the pixel space, the metric of nICA becomes more similar to
       the pixel basis and the performance with respect to coding
       efficiency can be seen to improved considerably.}
  \label{fig:RateDistortion}
\end{figure}

We compared the rate--distortion curves of wICA, nICA, wPCA and oPCA
(see Figure~\ref{fig:RateDistortion}).  Despite the fact that ICA is
optimal in terms of redundancy reduction (see
Table~\ref{tab:AverageLogLoss}), oPCA performs optimal with respect to coding
efficiency. wPCA in turn performes worst and remarkably similar to
wICA. Since wPCA and wICA differ only by an orthogonal transformation,
both representations are bound to the same metric.  oPCA is the only
transformation which has the same metric as the pixel representation
according to which the reconstruction error is determined. By
normalizing the length of the ICA basis vectors in the pixel space,
the metric of nICA becomes more similar to the pixel basis and the
performance with respect to coding efficiency improved
considerably. Nevertheless, for a fixed reconstrucion error the
discrete entropy after quantization in the oPCA basis is up to 1
bit/component {\it smaller} than for the corresponding nICA-basis.

In order to understand this result more precisely, we analyzed how the
quantization of the coefficients affects the two variables of the
rate--distortion function, {\em discrete entropy} and {\em reconstruction error}.

\begin{figure}[H]
  \begin{center}
    \includegraphics{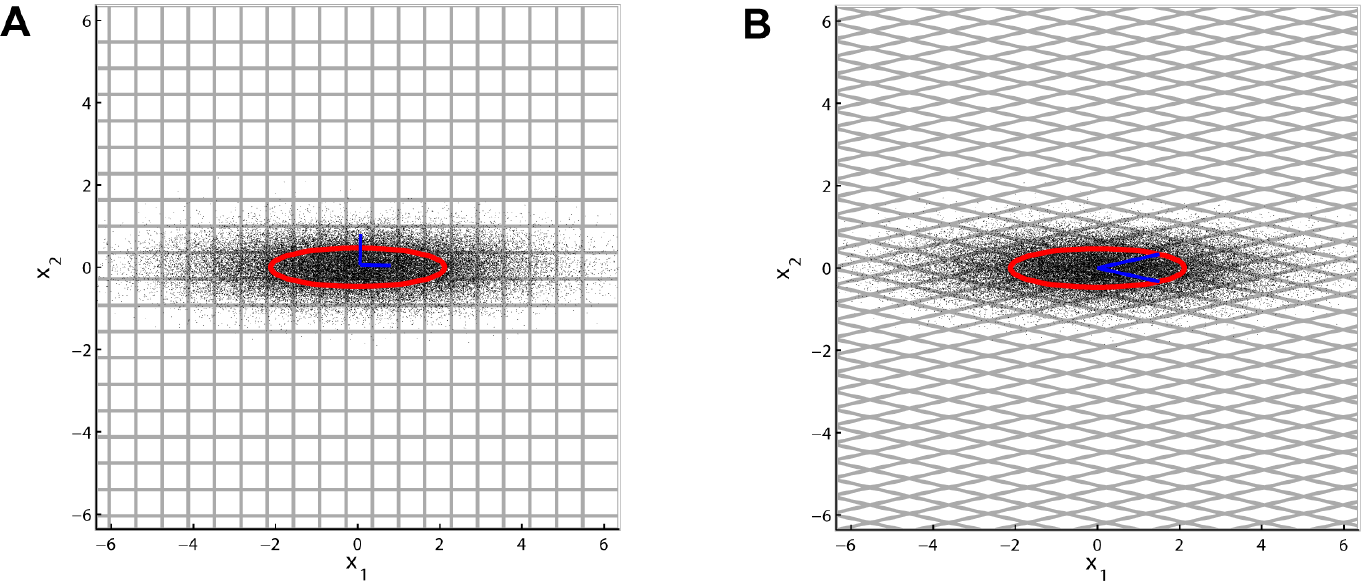}
  \end{center}
  \caption{{\bf The Partition Cell Shape is Crucial for the
      Quantization Error} \hspace{0.3cm} The quality of a source code
    depends on the shapes of the partition cells and on varying the
    sizes of the cells according to the source density.  When the
    cells are small (i.e. the entropy rate is high), then, the quality
    mainly depends on having cell shapes that minimize the average
    distance to the center of the cell. For a given volume, a body in
    Euclidean space that minimizes the average distance to the center
    is a sphere. The best packings (including the hexagonal case)
    cannot be achieved with linear transform codes.  Transform codes
    can only produce partitions into parallelepipeds, as shown here
    for two dimensions. The best parallelepipeds are cubes. Therefore
    PCA yields the (close to) optimal trade-off between minimizing the
    redundancy {\it and} the distortion as it is the only orthogonal
    decorrelation transform (see \cite{goyal:2001} for more details).
    The figure shows $50.000$ samples from a bivariate Gaussian random
    variable. Plot {\bf{\textsf{A}}} depicts a uniform binning (bin
    width $\Delta = 0.01$, only some bin borders are shown) induced by
    the only orthogonal basis for which the coefficients $x_1$ and
    $x_2$ are decorrelated. Plot {\bf{\textsf{B}}} shows uniform
    binning in a decorrelated, but not orthogonal basis (indicated by
    the blue lines). Both cases have been chosen such that the
    multi-information between the coefficients is identical and the
    same entropy rate was used to encode the signal. However, due to
    the shape of the bins in plot {\bf{\textsf{B}}} the total
    quadratic error increases from $0.4169$ to $0.9866$. The code for
    this example can be also downloaded from {\tt
      http://www.kyb.tuebingen.mpg.de/bethge/code/QICA/}.}
  \label{fig:cell_shape_rate_distortion}
\end{figure}

Figure~\ref{fig:cell_shape_rate_distortion} shows an illustrative
example in order to make the following analysis more intuitive.  The
example demonstrates that the quality of a transform code not only
depends on the redundancy of the coefficients but also on the shape of
the partition cells induced by the quantization. In particular, when
the cells are small (i.e. the entropy rate is high), then the
reconstruction error mainly depends on having cell shapes that
minimize the average distance to the center of the cell. Linear
transform codes can only produce partitions into parallelepipeds
(Figure~\ref{fig:cell_shape_rate_distortion}{\bf{\textsf B}}).  The
best parallelepipeds are cubes
(Figure~\ref{fig:cell_shape_rate_distortion}{\bf{\textsf A}}). This is
why PCA yields the (close to) optimal trade-off between minimizing the
redundancy {\it and} the distortion, as it is the only orthogonal
transform that yields uncorrelated coefficients.  For a more
comprehensive introduction to transform coding we refer the reader to
the excellent review by Goyal \cite{goyal:2001}.

\begin{figure}[H]

  \begin{center}
    \includegraphics{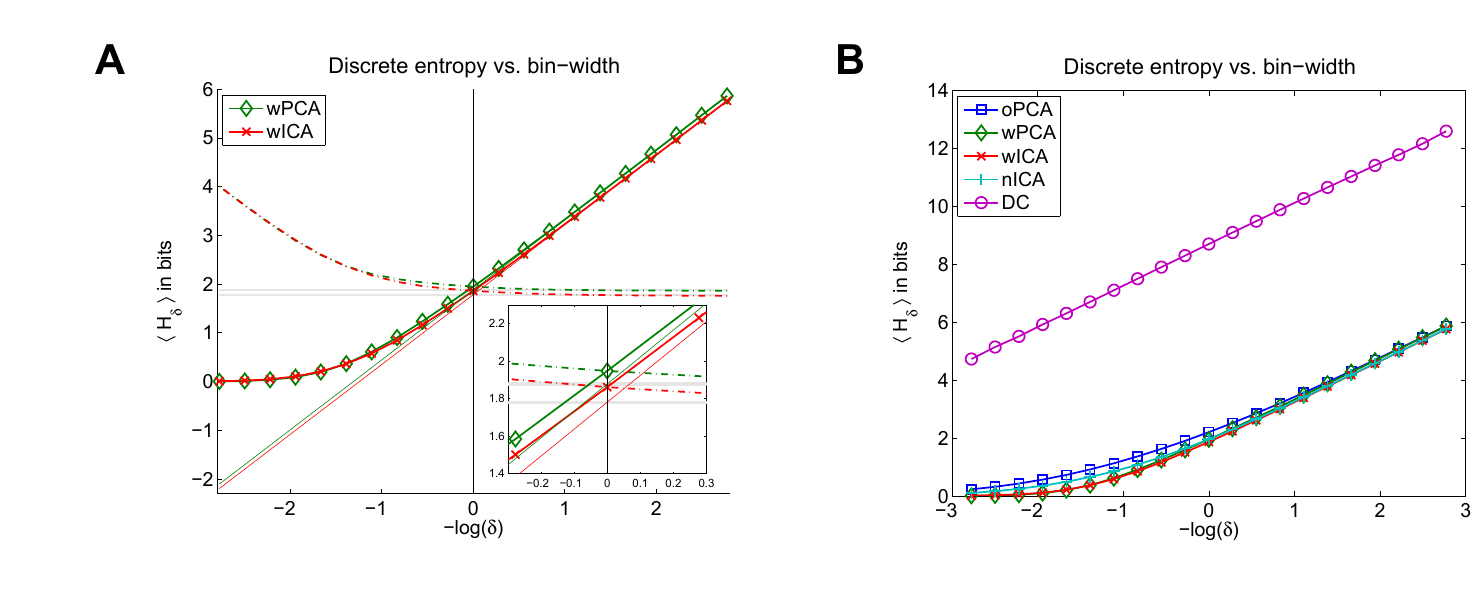}
  \end{center}
  \caption{{\bf Discrete vs. Differential Entropy } \hspace{0.3cm}
    {\bf\textsf{A.}} Relationship between discrete and differential
    entropy. Discrete entropy $\langle H_\delta \rangle$ averaged over
    all channels as a function of the negative log-bin-width.  The
    straight lines constitute the linear approximation to the
    asymptotic branch of the function.  Their interception with the
    y-axis are visualized by the gray shaded, horizontal lines.  The
    dashed lines represent $\langle h_{\delta} \rangle$ which converge
    to the gray shaded lines for $\delta \Ra 0$.  {\bf\textsf{B.}}
    There are only small differences in the average discrete entropy
    for oPCA, wPCA, wICA, nICA as a function of the negative
    log-bin-width. Since the discrete entropy of the DC component is
    the same for all transforms, it is not included in that average
    but plotted separately instead.}
  \label{fig:DiscreteEntropy}
\end{figure}

\paragraph{Discrete entropy}
Given a uniform binning of width $\delta$ the discrete entropy
$H_\delta$ of a probability density $p(x)$ is defined as
 \begin{equation}
   H_\delta = - \sum_i p_i \log p_i \quad \text{with} \quad p_i = \int_{B_i} p(x) dx \,,\\
 \end{equation}
 where $B_i$ denotes the interval defined by the $i$-th bin. For small
 bin-sizes $\delta \to 0$, there is a close relationship between {\it
   discrete} and {\it differential} entropy: Because of the mean value
 theorem we can approximate $p_i \approx p(\xi_i) \delta$ with $\xi
 \in B_i$, and hence

\begin{align*}
  H_\delta &\approx - \sum_i p(\xi_i) \delta \; \log[ p(\xi_i) \delta]\\
  &= \underbrace{- \sum_i \delta \; p(\xi_i) \log p(\xi_i)}_{
    \stackrel{\delta \Ra 0}{\Lra} -\int p(x) \log p(x)\; \text{d}x} -
  \log \delta \underbrace{\sum_i p(\xi_i) \delta}_{\stackrel{\delta
      \Ra 0}{\Lra}1} \, .
\end{align*}
Thus, we have the relationship $H_\delta \approx h - \log \delta$ for
sufficiently small $\delta$ (i.e. high-rate quantization). In other
words, $H_\delta$ asymptotically grows linearly with $(-\log \delta)$.
Therefore, we can fit a linear function to the asymptotic branch of
the function $H_\delta=H_\delta(-\log \delta)$ which is plotted in
Figure~\ref{fig:DiscreteEntropy}{\bf \textsf{A}} (more precisely we are plotting
the average over all dimensions). If we take the ordinate intercept of
the linear approximation, we obtain a nonparametric estimate of the
differential entropy which can be compared to the entropy estimates
reported above (Those estimates were determined with the OPT
  estimator).  Equivalently, one can consider the function
$h_\delta(-\log \delta) := H_\delta - (- \log\delta)$ which gives a
better visualization of the error of the linear approximation
(Figure~\ref{fig:DiscreteEntropy}, left, dashed line).  For
$h_\delta(-\log \delta)$ the differential entropy is obtained in the
limit $h=\lim_{(-\log \delta) \to \infty} h_\delta = \lim_{\delta \to
  0} h_\delta$.

This analysis shows that differences in differential entropy in fact
translate into differences in discrete entropy after uniform
quantization with sufficiently small bins. Accordingly, the
minimization of the multi-information as proposed by the redundancy
reduction hypothesis does in fact also minimize the discrete entropy
of a uniformly quantized code. In particular, if we look at the
discrete entropy of the four different transforms, oPCA, wPCA, wICA,
nICA (Figure~\ref{fig:DiscreteEntropy}{\bf \textsf{B}}), we find that
asymptotically the two PCA transforms require slightly more entropy
than the two ICA transforms, and there is no difference anymore
between oPCA and wPCA or wICA and nICA. This close relationship
between discrete and differential entropy for high-rate quantization,
however, is not sufficient to determine the coding efficiency
evaluated by the rate--distortion curve. The latter requires to
compare also the reconstruction error for the given quantization.

\paragraph{Reconstruction error}
The reconstruction error is defined as the mean squared distance in
the pixel basis between the original image and the image obtained by
reconstruction from the quantized coefficients of the considered
transformation. For the reconstruction, we simply use the inverse of
the considered transformation, which is optimal in the limit of
high-rate quantization.

\begin{figure}[H]
  \begin{center}
    \includegraphics[width=0.5\linewidth]{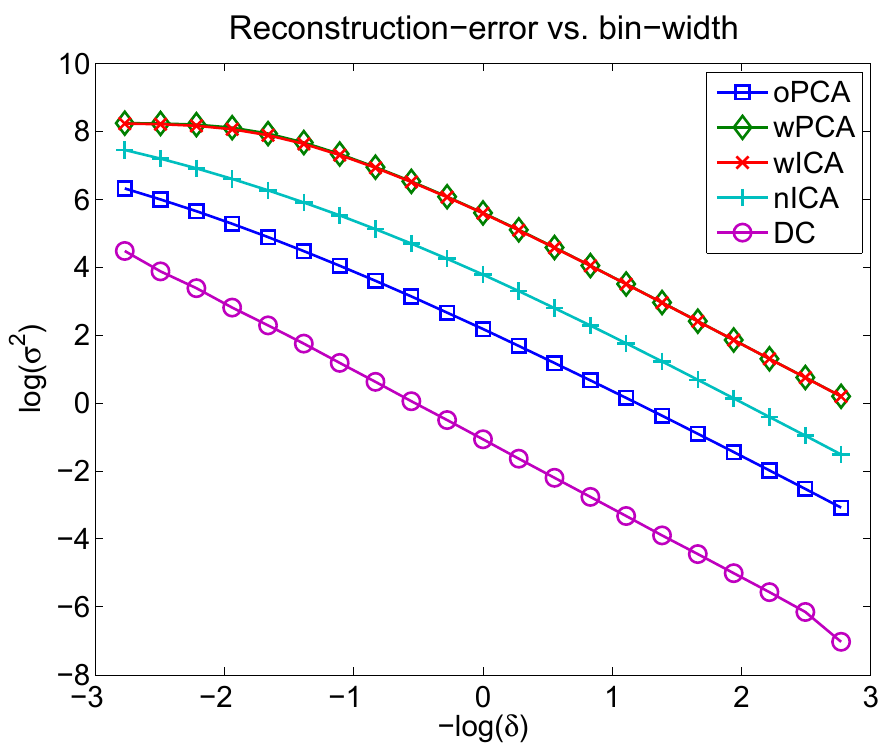}
  \end{center}
  \caption{{\bf Reconstruction Error vs. Bin Width of Discrete Entropy
    } \hspace{0.3cm} Reconstruction error $\sigma^2$ as a function of
    the bin-width $\delta$, shown on a logarithmic scale. The
    differences between the different transforms are relatively large.
    Only the two transformations with exactly the same metric, wPCA
    and wICA, exhibit no difference in the reconstruction error.  }
  \label{fig:ReconErr}
\end{figure}

When looking at the reconstruction error as a function of the
bin width (Figure~\ref{fig:ReconErr}) we can observe much more
pronounced differences between the different transformations than it
was the case for the entropy. As a consequence, the differences in the
reconstruction error turn out to be much more important for coding
efficiency than the differences in the entropy. Only the two
transformations with exactly the same metric, wPCA and wICA, exhibit
no difference in the reconstruction error. This suggests that
minimization of the multi-information is strictly related to efficient
coding if and only if the transformation with respect to the pixel
basis is orthogonal. As we have seen that the potential effect of
higher-order redundancy reduction is rather small, we expect that the
PCA transform constitutes a close approximation to the minimizer of
the multi-information among all orthogonal transforms because PCA is
the only orthogonal transform which removes all second-order
correlations.

\section*{Discussion}
\label{Sec:discussion}

The structural organization of orientation selectivity in
the primary visual cortex has been associated with self-organization
since the early seventies \citep{malsburg:1973}, and much progress has
been made to narrow down the range of possible models compatible with
the empirical findings
\citep[e.g.][]{kaschube:2006,wolf:2005,wimbauer:1997}. The link to
visual information processing, however, still remains elusive
\citep{horton:2005,olshausen:2005,masland:2007}.

More abstract unsupervised learning models which obtain orientation
selective filters using sparse coding \citep{olshausen:1996} or ICA
\citep{bell:1997} try to address this link between image processing
and the self-organization of neural structure. In particular, these
models not only seek to reproduce the orientation tuning properties of
V1 simple cells but they additionally address the question of how the
simple cell responses collectively can instantiate a representation
for arbitrary images. Furthermore, these image representations are
learned from an information theoretic principle assuming that the
learned filters exhibit advantageous coding properties.

The goal of this study is to quantitatively test how much evidence
there is for this assumption. To this end, we investigated three
criteria, the multi-information---i.e. the objective function of
ICA---, the average log-loss, and rate-distortion curves. There are a
number of previous studies which also tried to quantify how large the
advantage of the orientation selective ICA filters is relative to
second-order decorrelation transformations. In particular, four papers
\citep{lewicki:1999,wachtler:2001,LeeWacSej2002,bethge:2006}, are most
closely related to this study as all of them compare the average
log-loss of different transformations. Unfortunately, they do not
provide a coherent answer to the question how large the advantage of
ICA is compared to other decorrelation transforms.

Lewicki and Olshausen \citep{lewicki:1999} claim that their learned
bases show a 15 -- 20\% improvement over traditional bases, but it is
not really clear how exactly this number was obtained. According to
reported values in their study (Table 1 in \citep{lewicki:1999}) one
could conclude that the advantage of ICA over PCA is even larger.
Apart from that, a principle problem of this study is that the entire
analysis is based on a dataset in which all images have been
preprocessed with a bandpass filter as in \cite{olshausen:1996}. Since
bandpass filtering already removes a substantial fraction of
second-order correlations in natural images, their study is likely to
systematically underestimate the total amount of second-order
correlations in natural images. Therefore, a valid comparison between
second-order and higher-order redundancy reduction is not possible
from their study.

Lee {\it et al} \citep{wachtler:2001,LeeWacSej2002} reported an
advantage of over 100\% percent for ICA in the case of color images
and a more moderate but substantial gain of about 20\% for
gray-value images.  In order to avoid possible differences due to the
choice of data set we here used exactly the same data as in
\citep{wachtler:2001,LeeWacSej2002}.  Very consistently, we find only
a small advantage for ICA of less than five percent for both
multi-information and the average log-loss. In particular, we are not
able to reproduce the very large difference between color and
gray-value images that they reported.

Unfortunately, it is not clear which estimation procedure was used in
\citep{wachtler:2001,LeeWacSej2002}. Therefore, we cannot pinpoint
where the differences in the numbers ultimately come from. There seems
to be an inconsistency though between the kurtosis estimates and the
average log-loss they reported. The kurtosis they find for ICA is
smaller than twenty. Thus, for the exponential power family which fits
the coefficient statistics very well, we can derive that the Negentropy
has to be less than 0.52 bits per pixel (cf.\ Equations~(9) and (12) in
\citep{bethge:2006}). The Negentropy quantifies the difference between
the entropy of the maximum entropy distribution of same variance minus
the true entropy of the distribution. Since the redundancy reduction due
to PCA or any other decorrelation transform is around 4 bits per pixel,
the maximum possible advantage for ICA should be less than 15\% even
if one uses the kurtosis estimate taken from
\citep{wachtler:2001,LeeWacSej2002}.

The estimators for the measures used for comparing the different
linear transforms in the present study are well designed and have been
shown to give correct results on artificial data
\citep{bethge:2006}. Furthermore, Weiss and Freeman showed for
an undirected probabilistic image model that whitening already yields $98\%$
of the total performance \cite{weiss:2007}. Finally, the superior performance
of the simple SSD model with only two free parameters provides a very strong
explanation for why the gain achieved with ICA is so small relative to
a random decorrelation transform: Since a spherically symmetric model
is invariant under orthogonal transformations and provides a better
fit to the data, the actual shape of the filter does not seem to
be critical. It also shows that the fundamental assumption underlying
ICA---the data are well described by a linear generative model with
independent sources---is not justified in the case of natural images.

From all these results, we can safely conclude that the actual gain of
ICA compared to PCA is smaller than 5\% for both gray level images and
color images. 

\paragraph{Is smaller than 5\% really small?}
A valid question to ask is whether comparing the amount of
higher-order correlations to the amount of second-order correlations
is the right thing to do. The reason why ``bits'' are the
canonical units used to measure the multi-information and the average
log-loss, originates from a coding theoretical theorem which says that
the entropy is an (asymptotically) attainable lower bound to the average
code word length (Differential entropy can be related in a meaningful
way to discrete entropy assuming sufficiently fine quantization.) \cite{gray:1990}. 
Hence, if the brain uses codes which are close to
optimal, the multi-information measured in bits would correspond to
the amount of resources (e.g. number of neurons or number of spikes)
that it could save by the redundancy reduction. If the brain
uses a suboptimal code, say a grandmother code for instance, then the
number of necessary neurons would rather equal the number of states
which grows exponentially with entropy. If we now suppose that
$H_{PIX}>H_{RND}>H_{ICA}$ is the entropy of the pixels, of random
whitening, and of ICA after quantization, respectively, then

\begin{equation}
  \frac{H_{RND}-H_{ICA}}{H_{PIX}-H_{RND}} < 5\%
\end{equation}
as reported in Table~\ref{tab:MultiInfo}, implies that
\begin{equation}
  \frac{2^{H_{RND}}-2^{H_{ICA}}}{2^{H_{PIX}}-2^{H_{RND}}} << 5\% \, .
\end{equation}

In this sense we can argue that measuring bits (i.e. the ``log-number'' of
states) rather than the number of states is a conservative way of
showing that the improvement of ICA relative to decorrelation is
small. 

\begin{figure}[H]
  \begin{center}
    \includegraphics[width=1\linewidth]{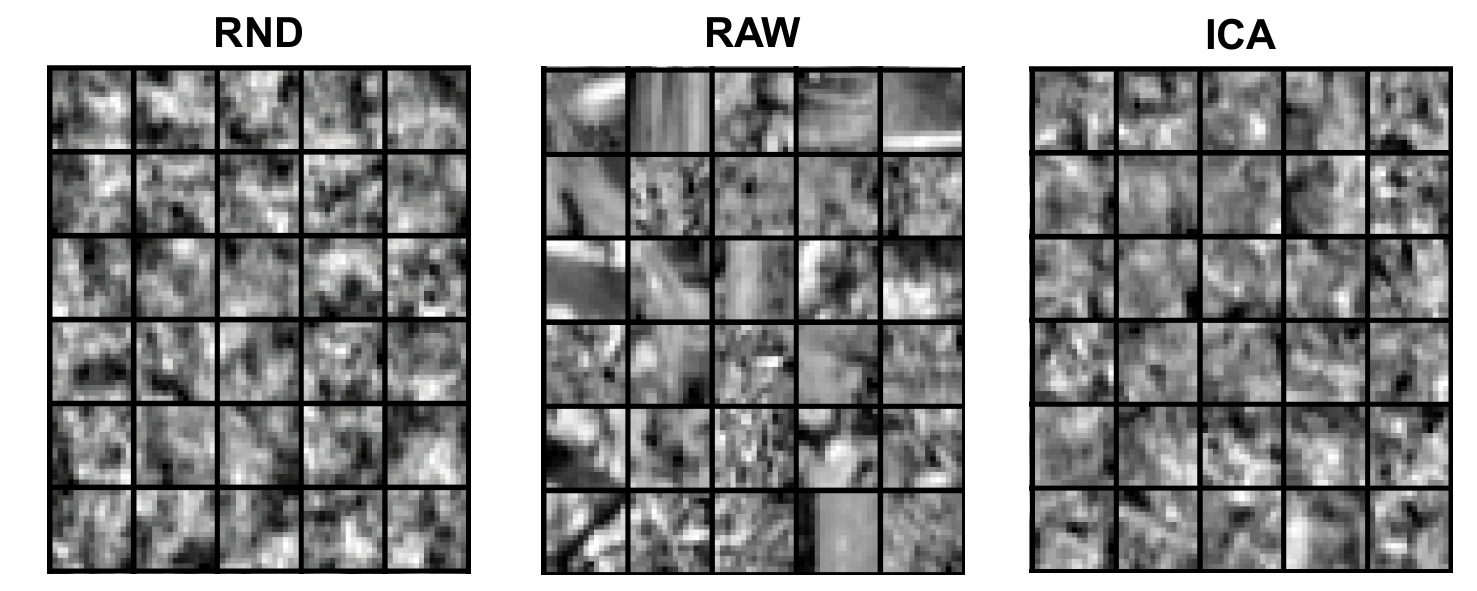}
  \end{center}
  \caption{{\bf Comparison of Patches Sampled From Different Image Models
    } \hspace{0.3cm} The figure demonstrates that the visual appearance of samples
    from the ICA image model (right) do not look significantly more similar to samples
    from natural images (middle) than samples from the image model (left) with random
    (pink noise) basis functions.}
  \label{fig:sampled_patches}
\end{figure}

We have computed complete rate distortion curves for ICA and PCA
showing that ICA performs even worse than PCA with respect to coding
efficiency because of its unfavorable metric.  By disentangling the
rate--distortion curves into {\it bin width---rate} curves and {\it
  bin width---distortion} curves, we show that the average log-loss
can only account for differences in the rate for a given quantization
bin width. It does not take into account, however, that the
reconstruction error of two transforms can be quite different for a
fixed bin width if they do not have the same metric, i.e. if they
differ by more than just an orthogonal transformation. This is due to
the fact that for sufficiently small bin widths the quality of the
encoding depends on having cell shapes that minimize the average
distance to the center of the cell \cite{goyal:2001}. Therefore,
non-orthogonal decorrelation transforms suffer from disadvantageous
cell shapes (Figure~\ref{fig:cell_shape_rate_distortion}). This
explains why ICA performs even worse than PCA with respect to coding
efficiency in terms of mean squared reconstruction error. Hence, the
advantage of ICA in redundancy reduction does not pay off in terms of
reconstruction error. Finally, also perceptually image patches sampled
from the ICA model do not look more similar to natural image patches
than those sampled from the random decorrelation basis
(Figure~\ref{fig:sampled_patches}).

In summary, we were not able thus far to come up with a meaningful
interpretation for which the improvement of ICA would be recognized as
being large. If such an interpretation exists it would very desirable
to know as it might point to a possible refinement of the
interpretation of the redundancy reduction hypothesis. On the basis of
the present study, however, it seems rather unlikely that such a
measure can be found. Instead, we think that our study provides solid
evidence that orientation selective filters in a {\it linear}
representation do not yield a clear advantage in terms of redundancy
reduction and are rather suboptimal in terms of coding efficiency.

\paragraph{What about nonparametric approaches?}
The focus on linear redundancy reduction models in this study is
motivated by the goal to first establish a solid and reproducible
result for the simplest possible case before moving on to more
involved nonlinear transformations.  Nevertheless, it is important to
discuss what we can expect if the restriction to linear
transformations is dropped. From a nonparametric analysis
\cite{petrov:2003}, Petrov and Zhaoping concluded that higher-order
correlations in general contribute only very little to the redundancy
in natural images and, hence, are probably not the main cause for the
receptive field properties in V1. The empirical support for this
claim, however, is not very strong as their comparison is based on mutual
information estimates within a very small neighborhood of five pixels
only. This is problematic as it is known that many kinds of
higher-order correlations in natural images manifest themselves only
in much higher-dimensional statistics \citep{bethge:2008}.
Furthermore, their estimate of the amount of second-order correlations
is not invariant against point-wise nonlinear transformations of the
pixel intensities.

In a more recent non-parametric study, Chandler and Field arrived at a
very different conclusion than Petrov and Zhaoping. They use
nearest-neighbor based methods to estimate the joint entropy of
natural images in comparison to pink noise and white noise
\cite{chandler:2007} where pink noise denotes Gaussian noise with the
same spectrum as that of natural images. As a result (Fig. 18) they find
a smaller difference between pink noise and white noise as between
natural images and pink noise. Thus, from their
finding, it seems that the amount of higher-order correlations in
natural images is even larger than the amount of second-order correlations.
Also this result has to be taken with care:
Apart from the fact that reliable non-parametric estimates in
high-dimensions are difficult to obtain even if one resorts to
nearest-neighbor based methods, the estimate of the amount of
second-order correlations in \cite{chandler:2007} is not invariant against pointwise nonlinear
transformations of the pixel intensities, too.

In summary,  the results of the present nonparametric studies do not yield a clear
picture how large the total amount of higher-order correlations in natural images really is.
In fact, estimating the absolute amount of multi-information is an extremely difficult task in
high dimensions. Therefore, the differences in the results can easily originate from
the different assumptions and approximations made in these studies. 

\paragraph{What about nonlinear image models?}
Apart from the non-parametric approaches, a large number of nonlinear
image models has been proposed over the years which are capable to
capture significantly more statistical regularities of natural images
than linear ICA can do (e.g. \cite{wainwright:2000, karklin:2003,
weiss:2007,Hyvaerinen:2007,osindero:2008,garrigues:2008,
seeger:2008,guerrero:2008,hammond:2008}).  In fact,
Olshausen and Field \cite{olshausen:1996} already used a more general
model than linear ICA when they originally derived the orientation
selective filters from higher-order redundancy reduction. In contrast
to plain ICA, they used an {\it overcomplete} generative model which
assumes more source signals than pixel dimensions. In addition, the
sources are modeled as latent variables like in a factor analysis
model.  That is the data is assumed to be generated according to
$\mathbf{x} = A\mathbf{s} + \xi$ where $A$ denotes the overcomplete dictionary,
$\mathbf{s}$ is distributed according to a sparse
factorial distribution, and $\xi$ is a Gaussian random variable.
The early quantitative study by Lewicki and Olshausen
\cite{lewicki:1999} could not demonstrate an advantage of overcomplete
coding in terms of coding efficiency and also the more recent work by
Seeger \cite{seeger:2008} seems to confirm this conclusion. The
addition of a Gaussian random variable $\xi$ to $A\mathbf{s}$, however is
likely to be advantageous as it may help to interpolate betweem the
plain ICA model on the one hand and the spherically symmetric model on
the other hand. A comparison of the average log-loss between this model
and plain ICA has not been done yet but we can expect that this model
can achieve a similar or even better match to the natural image statistics
as the spherically symmetric model.

The spherical symmetric model can also be modeled by a redundancy
reduction transformation which changes the radial component such that
the output distribution is sought to match a Gaussian distribution
\cite{lyu:2008}. Hence, the redundancy reduction of this model is very
similar to the average log-loss of the spherically symmetric
distribution.  From a biological vision point of view, this type of
model is particularly interesting as it allows one to draw a link to
divisive normalization, a prominent contrast gain control mechanism
observed for virtually all neurons in the early visual system. Our own
ongoing work \cite{sinz:2008} shows that this idea can be
generalized to a larger class of $L_p$-spherically symmetric
distributions \cite{Hyvaerinen:2007}. In this way, it is possible to
find an optimal interpolation between ICA and the spherically
symmetric case \cite{sinz:2008a}. That is, one can combine
orientation selectivity with divisive normalization in a joint model.
Our preliminary results suggests that optimal divisive normalization
together with orientation selectivity allows for about 10\%
improvement while divisive normalization alone (i.e. the spherical
symmetric model) is only 2\% worse \cite{sinz:2008}.

Note that, in general, orientation selectivity can only play a critical
role if one assumes some form of regularization on the
function space of possible transformations. Otherwise, the optimization
problem would be highly degenerated \cite{hyvarinen:1999}. Thus,
even if we drop the constraint of linearity, we probably want to stay
somewhat close to a linear function space. Typical examples would be
the generalized linear model (i.e. adding pointwise nonlinearities)
which does not have an effect on the multi-information at all, or---more
interestingly---a scaling nonlinearity such as divisive normalization
which we discussed above.

\paragraph{Concluding remarks}
Taken together, the current stage of statistical image modeling and
quantitative model comparison suggests that it is rather unlikely that orientation 
selectivity plays a critical role for redundancy reduction even if the class of 
transformations is not limited to linear ones. 
This finding urges us to rethink how we can interpret the function of
orientation selectivity in primary visual cortex. For one, we should
further explore and quantify how nonlinearities and other known properties of
V1 contribute to redundancy reduction.  On the other hand, we should also
try to extend the quantitative approach to include other goals than
redundancy reduction only. While there are certainly many ideas about
possible functions of orientation selective filtering, the crucial
challenge is to turn those ideas into a quantitatively testable
framework. For instance, many algorithms for edge detection, image
segmentation, and texture classification use orientation selective
filtering as a preprocessing step. Most frequently, however, it is not
clear how critical this choice of image basis is for solving these
tasks.

In order to achieve a more principled take on the potential use
of orientation selectivity for early vision we are currently trying
to extend the efficient coding framework to deal with other loss
functions than mean squared error on the image pixels. Obviously,
the goal of the visual system is not to preserve the representation
of the visual input. Instead, seeing means to make successful
predictions about behaviorally relevant aspects of the environment
\cite{helmholtz:1878}. Since 3D shape inference is necessary to
almost any naturally relevant task, it seems particularly interesting
to explore the role of orientation selectivity in the context of 3D
shape inference \cite{fleming:2004}. For a quantitative account
of this problem one can seek to minimize the reconstruction error
for the 3D shape rather than for its 2D image. Certainly, this task
is much more involved than image reconstruction and there is still a
long way to go. Nevertheless, we believe that going in this direction
is  a worthwhile enterprise which eventually may help us to unravel
the principles of neural processing  in the brain that are ultimately
responsible for our ability to see.

\section*{Acknowledgements}
We would like to thank Philipp Berens, Roland Fleming and Jakob Macke for helpful
comments on the manuscript. This work is supported by the German
Ministry of Education, Science, Research and Technology through
the Bernstein prize to MB (BMBF; FKZ: 01GQ0601), a scholarship to FS
by the German National Academic Foundation, and the Max
Planck Society. 



\end{document}

%% file: definitions.tex
\newcommand{\expc}[2][]{{\displaystyle\mathop{\mathbb{E}}_{#1}\left[ #2\right]}}
\newcommand{\expe}[2][]{{\displaystyle\mathop{\mathbb{\hat E}}_{#1}\left[ #2\right]}}

\newcommand{\MB}[1]{\mathbf{#1}}

\newcommand{\Ra}{\rightarrow}
\newcommand{\Lra}{\longrightarrow}

%% file: Eichhorn_et_al_2008_arXiv.bbl
\begin{thebibliography}{10}
\providecommand{\url}[1]{\texttt{#1}}
\providecommand{\urlprefix}{URL }
\expandafter\ifx\csname urlstyle\endcsname\relax
  \providecommand{\doi}[1]{doi:\discretionary{}{}{}#1}\else
  \providecommand{\doi}{doi:\discretionary{}{}{}\begingroup
  \urlstyle{rm}\Url}\fi
\providecommand{\bibAnnoteFile}[1]{%
  \IfFileExists{#1}{\begin{quotation}\noindent\textsc{Key:} #1\\
  \textsc{Annotation:}\ \input{#1}\end{quotation}}{}}
\providecommand{\bibAnnote}[2]{%
  \begin{quotation}\noindent\textsc{Key:} #1\\
  \textsc{Annotation:}\ #2\end{quotation}}
\providecommand{\eprint}[2][]{\url{#2}}

\bibitem{attneave:1954}
Attneave F (1954) Informational aspects of visual perception.
\newblock Psychological Review 61:183--193.
\bibAnnoteFile{attneave:1954}

\bibitem{barlow:1959}
Barlow H (1959) Sensory mechanisms, the reduction of redundancy, and
  intelligence.
\newblock In: The Mechanisation of Thought Processes. London: Her Majesty's
  Stationery Office, pp. 535--539.
\bibAnnoteFile{barlow:1959}

\bibitem{atick:1992}
Atick J (1992) Could information theory provide an ecological theory of sensory
  processing?
\newblock Network 3:213--251.
\bibAnnoteFile{atick:1992}

\bibitem{barlow:1989}
Barlow H (1989) Unsupervised learning.
\newblock Neural Comput 1:295--311.
\bibAnnoteFile{barlow:1989}

\bibitem{watanabe:1981}
Watanabe S (1981) Pattern recognition as a quest for minimum entropy.
\newblock Pattern Recognition 13:381--387.
\bibAnnoteFile{watanabe:1981}

\bibitem{foldiak:1990}
F{\"o}ldi{\'a}k (1990) Forming sparse representations by local anti-hebbian
  learning.
\newblock Biological Cybernetics 64:165--170.
\bibAnnoteFile{foldiak:1990}

\bibitem{olshausen:1996}
Olshausen B, Field D (1996) Emergence of simple-cell receptive field properties
  by learning a sparse code for natural images.
\newblock Nature 381:560--561.
\bibAnnoteFile{olshausen:1996}

\bibitem{foldiak:1991}
F\"oldiak P (1991) Learning invariance from transformation sequences.
\newblock Neural Comput 3:194--200.
\bibAnnoteFile{foldiak:1991}

\bibitem{bialek:2001}
Bialek W, Nemenman I, Tishby N (2001) Predictability, complexity, and learning.
\newblock Neural Computation 13:2409–2463.
\bibAnnoteFile{bialek:2001}

\bibitem{becker:1992}
Becker S, Hinton GE (1992) {Self-organizing neural network that discovers
  surfaces in random-dot stereograms}.
\newblock Nature 355:161--163.
\bibAnnoteFile{becker:1992}

\bibitem{Friedman:1984}
Friedman JH, Stuetzle W, Schroeder A (1984) Projection pursuit density
  estimation.
\newblock Journal of the American Statistical Association 19:599--608.
\bibAnnoteFile{Friedman:1984}

\bibitem{simoncelli:2001}
Simoncelli E, Olshausen B (2001) Natural image statistics and neural
  representation.
\newblock Ann Rev Neurosci 24:1193--1216.
\bibAnnoteFile{simoncelli:2001}

\bibitem{li:2006}
Zhaoping L (2006) Theoretical understanding of the early visual processes by
  data compression and data selection.
\newblock Network (Bristol, England) 17:301--34.
\newblock PMID: 17283516.
\bibAnnoteFile{li:2006}

\bibitem{buchsbaum:1983}
Buchsbaum G, Gottschalk A (1983) Trichromacy, opponent colours coding and
  optimum colour information transmission in the retina.
\newblock Proceedings of the Royal Society of London Series B, Biological
  Sciences 220:89--113.
\bibAnnoteFile{buchsbaum:1983}

\bibitem{atick:1992a}
Atick J, Redlich A (1992) What does the retina know about natural scenes.
\newblock Neural Computation 4:196--210.
\bibAnnoteFile{atick:1992a}

\bibitem{vanHateren:1993}
van Hateren J (1993) Spatiotemporal contrast sensitivity of early vision.
\newblock Vision research 33:257--67.
\bibAnnoteFile{vanHateren:1993}

\bibitem{dong:1995}
Dong DW, Atick JJ (1995) Statistics of natural time-varying images.
\newblock Network: Computation in Neural Systems 6:345--358.
\bibAnnoteFile{dong:1995}

\bibitem{dan:1996}
Dan Y, Atick JJ, Reid RC (1996) {Efficient coding of natural scenes in the
  lateral geniculate nucleus: experimental test of a computational theory}.
\newblock J Neurosci 16:3351--3362.
\bibAnnoteFile{dan:1996}

\bibitem{hancock:1992}
Hancock PJB, Baddeley RJ, Smith LS (1992) The principal components of natural
  images.
\newblock Network 3:61--70.
\bibAnnoteFile{hancock:1992}

\bibitem{li:1994}
Li Z, Atick JJ (1994) Toward a theory of the striate cortex.
\newblock Neural Comput 6:127--146.
\bibAnnoteFile{li:1994}

\bibitem{bell:1997}
Bell A, Sejnowski T (1997) The ``independent components'' of natural scenes are
  edge filters.
\newblock Vision Res 37:3327--38.
\bibAnnoteFile{bell:1997}

\bibitem{lewicki:1999}
Lewicki M, Olshausen B (1999) Probabilistic framework for the adaptation and
  comparison of image codes.
\newblock J Opt Soc Am A 16:1587--1601.
\bibAnnoteFile{lewicki:1999}

\bibitem{wachtler:2001}
Wachtler T, Lee TW, Sejnowski TJ (2001) Chromatic structure of natural scenes.
\newblock Journal of the Optical Society of America A, Optics, image science,
  and vision 18:65--77.
\newblock PMID: 11152005.
\bibAnnoteFile{wachtler:2001}

\bibitem{LeeWacSej2002}
Lee TW, Wachtler T, Sejnowski TJ (2002) Color opponency is an efficient
  representation of spectral properties in natural scenes.
\newblock Vision Res 42:2095--2103.
\bibAnnoteFile{LeeWacSej2002}

\bibitem{petrov:2003}
Y P, Zhaoping L (2003) Local correlations, information redundancy, and the
  sufficient pixel depth in natural images.
\newblock Journal of Optical Society of America 20:56--66.
\bibAnnoteFile{petrov:2003}

\bibitem{bethge:2006}
Bethge M (2006) Factorial coding of natural images: How effective are linear
  model in removing higher-order dependencies?
\newblock J Opt Soc Am A 23:1253--1268.
\bibAnnoteFile{bethge:2006}

\bibitem{perez:1977}
Perez A (1977) $\varepsilon$-admissible simplification of the dependence
  structure of a set of random variables.
\newblock Kybernetika 13:439--44.
\bibAnnoteFile{perez:1977}

\bibitem{cover:1991}
Cover T, Thomas J (1991) Elements of information theory.
\newblock New York: J. Wiley \& Sons.
\bibAnnoteFile{cover:1991}

\bibitem{bernardo:1979}
Bernardo JM (1979) Expected information as expected utility.
\newblock The Annals of Statistics 7:686--690.
\bibAnnoteFile{bernardo:1979}

\bibitem{lewicki:2000}
Lewicki M, Sejnowski T (2000) Learning overcomplete representations.
\newblock Neural Comput 12:337--365.
\bibAnnoteFile{lewicki:2000}

\bibitem{vanHulle:2005}
Hulle MMV (2005) Mixture density modeling, kullback-leibler divergence, and
  differential log-likelihood.
\newblock Signal Processing 85:951--963.
\bibAnnoteFile{vanHulle:2005}

\bibitem{nadal:1994}
Nadal J, Parga N (1994) Nonlinear neurons in the low-noise limit: a factorial
  code maximizes information transfer.
\newblock Network: Comput Neural Syst 5:565--581.
\bibAnnoteFile{nadal:1994}

\bibitem{bell:1995}
Bell A, Sejnowski T (1995) An information maximisation approach to blind
  separation and blind deconvolution.
\newblock Neural Computation 7:1129--59.
\bibAnnoteFile{bell:1995}

\bibitem{goyal:2001}
Goyal V (2001) {Theoretical Foundations of Transform Coding}.
\newblock IEEE Signal Processing Magazine 18:9--21.
\bibAnnoteFile{goyal:2001}

\bibitem{gray:1990}
Gray R (1990) Entropy and Information Theory.
\newblock New York: Springer.
\bibAnnoteFile{gray:1990}

\bibitem{wang:2002}
Wang Z, Bovic A, Lu L (2002) Why is image quality assessment so difficult?
\newblock In: Acoustics, Speech, and Signal Processing, 2002. Proceedings.
  (ICASSP '02). IEEE International Conference on. volume~4, pp. 3313--3316.
\bibAnnoteFile{wang:2002}

\bibitem{gray:1998}
Gray R, Neuhoff D (1998) Quantization.
\newblock Information Theory, IEEE Transactions on 44:2325--2383.
\bibAnnoteFile{gray:1998}

\bibitem{gish:1968}
Gish H, Pierce JN (1968) Asymptotically efficient quantizing.
\newblock IEEE Trans Inform Theory 14:676--683.
\bibAnnoteFile{gish:1968}

\bibitem{fan:1955}
Fan K, Hoffman AJ (1955) Some metric inequalities in the space of matrices.
\newblock roceedings of the American Mathematical Society 6:111--116.
\bibAnnoteFile{fan:1955}

\bibitem{srivastava:2003}
Srivastava A, Lee A, Simoncelli E, Zhu S (2003) On advances in statistical
  modeling of natural images.
\newblock Journal of Mathematical Imaging and Vision 18:17--33.
\bibAnnoteFile{srivastava:2003}

\bibitem{hyvarinen_book:2001}
Hyv\"arinen A, Karhunen J, Oja E (2001) Independent Component Analysis.
\newblock John Wiley \& Sons.
\bibAnnoteFile{hyvarinen_book:2001}

\bibitem{EdeAriSmi1999}
Edelman A, Arias TA, Smith ST (1999) The geometry of algorithms with
  orthogonality constraints.
\newblock SIAM J Matrix Anal Appl 20:303--353.
\bibAnnoteFile{EdeAriSmi1999}

\bibitem{NumRecC1992}
Press WH, Teukolsky SA, Vetterling WT, Flannery BP (1992) Numerical Recipes in
  C: The Art of Scientific Computing.
\newblock New York, NY, USA: Cambridge University Press.
\bibAnnoteFile{NumRecC1992}

\bibitem{maxwell:1855}
Maxwell JC (1855) {Experiments on colour as perceived by the eye, with remarks
  on colour- blindness}.
\newblock Transactions of the Royal Society of Edinburgh XXI 2:275--298.
\bibAnnoteFile{maxwell:1855}

\bibitem{zetzsche:1999}
Zetzsche C, Krieger G, , Wegmann B (1999) The atoms of vision: Cartesian or
  polar?
\newblock Journals of the Optical Society of America 16:1554--1565.
\bibAnnoteFile{zetzsche:1999}

\bibitem{lyu:2008}
Lyu S, Simoncelli EP (2008) Nonlinear image representation using divisive
  normalization.
\newblock In: Proc. Computer Vision and Pattern Recognition.
\newblock To Appear.
\bibAnnoteFile{lyu:2008}

\bibitem{brelstaff:1995}
Brelstaff GJ, Parraga A, Troscianko T, Carr D (1995) Hyperspectral camera
  system: acquisition and analysis.
\newblock In: Lurie BJ, Pearson JJ, Zilioli E, editors, Proceedings of SPIE.
  volume 2587, pp. 150--159.
\newblock The database can be downloaded from:
  \url{http://psy223.psy.bris.ac.uk/hyper/}.
\bibAnnoteFile{brelstaff:1995}

\bibitem{baddeley:1996}
Baddeley R (1996) An effeicient code in v1.
\newblock Nature 381:560--561.
\bibAnnoteFile{baddeley:1996}

\bibitem{malsburg:1973}
Malsburg (1973) Self-organization of orientation sensitive cells in the striate
  cortex.
\newblock Biological Cybernetics 14:85--100.
\bibAnnoteFile{malsburg:1973}

\bibitem{kaschube:2006}
Kaschube M, Schnabel M, Loewel S, Coppola D, White LE, et~al. (2006) Universal
  pinwheel statistics in the visual cortex.
\newblock In: Neuroscience Meeting Planner. 545.9/T11, Atlanta, GA: Society for
  Neuroscience.
\bibAnnoteFile{kaschube:2006}

\bibitem{wolf:2005}
Wolf F (2005) Symmetry, multistability, and long-range interactions in brain
  development.
\newblock Physical Review Letters 95:208701--4.
\bibAnnoteFile{wolf:2005}

\bibitem{wimbauer:1997}
Wimbauer, Wenisch, Miller, van Hemmen (1997) Development of spatiotemporal
  receptive fields of simple cells: I. model formulation.
\newblock Biological Cybernetics 77:453--461.
\bibAnnoteFile{wimbauer:1997}

\bibitem{horton:2005}
Horton JC, Adams DL (2005) The cortical column: a structure without a function.
\newblock Philosophical transactions of the Royal Society of London Series B,
  Biological sciences 360:837--62.
\newblock PMID: 15937015.
\bibAnnoteFile{horton:2005}

\bibitem{olshausen:2005}
Olshausen BA, Field DJ (2005) {How close are we to understanding v1?}
\newblock Neural Comput 17:1665--1699.
\bibAnnoteFile{olshausen:2005}

\bibitem{masland:2007}
Masland RH, Martin PR (2007) The unsolved mystery of vision.
\newblock Current biology : CB 17:R577--82.
\newblock PMID: 17686423.
\bibAnnoteFile{masland:2007}

\bibitem{weiss:2007}
Weiss Y, Freeman W (2007) What makes a good model of natural images?
\newblock Computer Vision and Pattern Recognition, 2007 CVPR '07 IEEE
  Conference on :1--8\doi{10.1109/CVPR.2007.383092}.
\bibAnnoteFile{weiss:2007}

\bibitem{bethge:2008}
Bethge M, Berens P (2008) Near-maximum entropy models for binary neural
  representations of natural images.
\newblock In: Platt J, Koller D, Singer Y, Roweis S, editors, Advances in
  Neural Information Processing Systems 20, Cambridge, MA: MIT Press. pp.
  97--104.
\bibAnnoteFile{bethge:2008}

\bibitem{chandler:2007}
Chandler DM, Field DJ (2007) Estimates of the information content and
  dimensionality of natural scenes from proximity distributions.
\newblock J Opt Soc Am A 24:922--941.
\bibAnnoteFile{chandler:2007}

\bibitem{wainwright:2000}
Wainwright M, Simoncelli E (2000) Scale mixtures of {Gaussians} and the
  statistics of natural images.
\newblock In: Solla S, Leen T, {M\"{u}ller} KR, editors, Adv. Neural
  Information Processing Systems (NIPS*99). Cambridge, MA: MIT Press,
  volume~12, pp. 855--861.
\bibAnnoteFile{wainwright:2000}

\bibitem{karklin:2003}
Karklin Y, Lewicki M (2003) {Learning higher-order structures in natural
  images}.
\newblock Network 14:483--499.
\bibAnnoteFile{karklin:2003}

\bibitem{Hyvaerinen:2007}
Hyv{\"a}rinen A, K{\"o}ster U (2007) Complex cell pooling and the statistics of
  natural images.
\newblock Network 18:81--100.
\bibAnnoteFile{Hyvaerinen:2007}

\bibitem{osindero:2008}
Osindero S, Hinton G (2008) Modeling image patches with a directed hierarchy of
  markov random fields.
\newblock In: Platt J, Koller D, Singer Y, Roweis S, editors, Advances in
  Neural Information Processing Systems 20, Cambridge, MA: MIT Press. pp.
  1121--1128.
\bibAnnoteFile{osindero:2008}

\bibitem{garrigues:2008}
Garrigues P, Olshausen B (2008) Learning horizontal connections in a sparse
  coding model of natural images.
\newblock In: Platt J, Koller D, Singer Y, Roweis S, editors, Advances in
  Neural Information Processing Systems 20, Cambridge, MA: MIT Press. pp.
  505--512.
\bibAnnoteFile{garrigues:2008}

\bibitem{seeger:2008}
Seeger MW (2008) Bayesian inference and optimal design for the sparse linear
  model.
\newblock Journal of Machine Learning Research 9:759--813.
\newblock
  \urlprefix\url{http://www.jmlr.org/papers/volume9/seeger08a/seeger08a.pdf}.
\bibAnnoteFile{seeger:2008}

\bibitem{guerrero:2008}
Guerrero-Col\'on JA, Simoncelli EP, Portilla J (2008) Image denoising using
  mixtures of {Gaussian} scale mixtures.
\newblock In: Proc 15th IEEE Int'l Conf on Image Proc. San Diego, CA: IEEE
  Computer Society.
\bibAnnoteFile{guerrero:2008}

\bibitem{hammond:2008}
Hammond DK, Simoncelli EP (2008) Image modeling and denoising with
  orientation-adapted {Gaussian} scale mixtures.
\newblock IEEE Trans Image Processing Accepted for publication.
\bibAnnoteFile{hammond:2008}

\bibitem{sinz:2008}
Sinz FH, Bethge M (2008) How much can orientation selectivity and contrast gain
  control reduce the redundancies in natural images.
\newblock Technical Report 169.
\bibAnnoteFile{sinz:2008}

\bibitem{sinz:2008a}
Sinz FH, Gerwinn S, Bethge M (2008) Characterization of the p-generalized
  normal distribution.
\newblock Journal of Multivariate Analysis .
\bibAnnoteFile{sinz:2008a}

\bibitem{hyvarinen:1999}
Hyv\"arinen A (1999) Survey on independent component analysis.
\newblock Neural Computing Surveys 2:94--128.
\bibAnnoteFile{hyvarinen:1999}

\bibitem{helmholtz:1878}
Helmholtz H (1878) The facts of perception.
\newblock In: Kahl R, editor, Selected Writings of Hermann Helmholtz,
  Middletown, CT: Wesleyan University Press.
\bibAnnoteFile{helmholtz:1878}

\bibitem{fleming:2004}
Fleming RW, Torralba A, Adelson EH (2004) {Specular reflections and the
  perception of shape}.
\newblock J Vis 4:798--820.
\newblock \urlprefix\url{http://journalofvision.org/4/9/10/}.
\newblock
  \eprint{http://journalofvision.org/4/9/10/Fleming-2004-jov-4-9-10.pdf}.
\bibAnnoteFile{fleming:2004}

\end{thebibliography}
